# The Nature of the $H_2$-Emitting Gas in the Crab Nebula[*]

C. T. Richardson[1†], J. A. Baldwin[1,2†], G. J. Ferland[3†], E. D. Loh[1†], C. A. Kuehn[1†], A. C. Fabian[4†] & Philippe Salomé[5†]

[1]Department of Physics and Astronomy, Michigan State University, East Lansing, MI 48824-2320 USA
[2]Visiting Astronomer, Kitt Peak National Observatory, National Optical Astronomy Observatory, which is operated by the Association of Universities for Research in Astronomy (AURA) under cooperative agreement with the National Science Foundation.
[3]Department of Physics, University of Kentucky, Lexington, KY 40506, USA
[4]Institute of Astronomy, University of Cambridge, Madingley Road, Cambridge CB3 0HA
[5]LERMA & UMR8112 du CNRS, Observatoire de Paris, 61 Av. de l'Observatoire, F-75014 Paris, France

**Abstract**

Understanding how molecules and dust might have formed within a rapidly expanding young supernova remnant is important because of the obvious application to vigorous supernova activity at very high redshift. In previous papers we have mapped the Crab Nebula in a roto-vibrational $H_2$ emission line, and then measured the molecular excitation temperature for a few of the brighter $H_2$-emitting knots that we have found to be scattered throughout the Crab Nebula's filaments. We found that $H_2$ emission is often quite strong, correlates with optical low-ionization emission lines, and has a surprisingly high excitation temperature. Here we study Knot 51, a representative, bright example. It is a spatially isolated structure for which we have available long slit optical and NIR spectra covering emission lines from ionized, neutral, and molecular gas, as well as HST visible and SOAR Telescope NIR narrow-band images. We present a series of CLOUDY simulations to probe the excitation mechanisms, formation processes and dust content in environments that can produce the observed $H_2$ emission. There is still considerable ambiguity about the geometry of Knot 51, so we do not try for an exact match between model and observations. Rather, we aim to explain how the bright $H_2$ emission lines can be formed from within a cloud the size of Knot 51 that also produces the observed optical emission from ionized and neutral gas. Our models that are powered only by the Crab's synchrotron radiation are ruled out because they are not able to reproduce the observed strong $H_2$ emission coming from thermally populated levels. The simulations that come closest to fitting the observations (although they still have conspicuous discrepancies) have the core of Knot 51 almost entirely atomic with the $H_2$ emission coming from just a trace molecular component, and in which there is extra heating. In this unusual environment, $H_2$ forms primarily through $H^-$ by associative detachment rather than by grain catalysis. In this picture, the 55 $H_2$-emitting cores that we have previously catalogued in the Crab have a total mass of about 0.1 M_sun, which is about 5% of the total mass of the system of filaments. We also explore the effect of varying the dust abundance. We discuss possible future observations that could further elucidate the nature of these $H_2$ knots.

## 1. Introduction

The Crab Nebula (hereafter "the Crab") presents a unique opportunity to study not only the properties and evolution of a very young supernova remnant which is not yet interacting with the ISM, but also in a more general way the details of the physical processes that occur in filamentary condensations exposed to a harsh environment of high-energy photons and particles. The relative proximity of the Crab, at a distance $d$ ~2.0 kpc (Trimble 1968), and the well-known date of the explosion, 1054 A.D. (Duyvendak 1942; Mayall & Oort 1942), allow detailed observations and constraints. The Crab filaments have been the subject of many previous investigations (see the review articles by Davidson & Fesen 1985 and by Hester 2008). The Crab's synchrotron radiation field has been taken into account as the energy source for the emitting regions, but not the effects of the high-energy particles that must also be present. Most of this previous work has concentrated on the ionized gas, generally with the aim of using optical emission lines to measure the chemical abundances in the aftermath of the supernova explosion. $H_2$ molecules are also present in the Crab, as was shown by the pioneering infrared observations and spectroscopy by Graham, Wright & Longmore (1990; hereafter G90). Those measurements were made with quite low (20" FWHM) angular resolution in just three locations.

We have been carrying out a much more detailed study of the molecular content of the Crab, mapping most of the nebula at sub-arcsec angular resolution in the NIR $H_2$ 2.12 μm emission line, and tying those results together with both existing and new spectra and images over a wide range of wavelengths. In our first paper in this series (Loh, Baldwin & Ferland 2010; hereafter Paper I), we investigated the brightest $H_2$ feature, Knot 1, for which the molecular content was estimated to be $M_{mol}$ ~ 5x10$^{-5}$ M_sun. We arrived at this estimate under the assumption that the $H_2$ emission is collisionally excited in a predominately molecular core. Then, in Loh et al. (2011) (hereafter Paper II) we presented our full near-infrared imaging survey made with the Spartan Infrared Camera on the 4 m SOAR telescope. We detected 55 compact (1-2" diameter) knots, spread across much of the face of the Crab, that radiate strongly in the $H_2$ 2.12 μm line. We showed that the positions of these knots correlate strongly with compact, low-ionization regions that are conspicuous in Hubble Space Telescope (HST) [S II] λλ6720 images; [S II] and also [O I] are $H_2$ tracers. The low-ionization regions are found on or next to the bright filaments seen in higher ionization lines such as [O III] λ5007. We used optical spectra to measure radial velocities of most of the [S II] features. Then in Paper III in this series (Loh et al. 2012), we used K-band spectra to measure the $H_2$ excitation temperature of 7 of the brightest knots. From the ratio of the S(1) 2-1 and S(0) 1-0 lines we found $T_{exc}$ ~ 2000-3000 K, which we argue is also the kinetic gas temperature in the regions where $H_2$ lines form under the conditions present in the Crab (Ferland 2011). These temperatures are just below the dissociation temperature of $H_2$.

There is also dust in the Crab, as can be seen both from the weak thermal dust emission peak at ~ 80 μm in the Crab's continuum spectrum (Marsden et al. 1984), and from the presence of many small dust globules seen in silhouette against the Crab's synchrotron continuum (Fesen & Blair 1990, Hester et al. 1990). The mass of dust in the filaments is uncertain, ranging from $10^{-3}$ M_sun (Temim et al 2006, 2012; Hester 2008) to 0.2 M_sun (Gomez et al. 2012). However, the gas to dust ratio could vary considerably between the filament cores and the ionized gas, and different regions could have both different abundances and different dust-to-gas ratios. Sankrit et al. (2008) estimate a dust-to-gas mass ratio in the cores that is a factor of 10 above the ISM value.

The ionized, neutral and molecular gas and the dust are distributed throughout a complex system of filaments that extend down into the synchrotron plasma and hence must be bombarded from all sides by intense synchrotron radiation and also by the high-energy particles that emit the synchrotron radiation. A more complete understanding of the nature of these filaments and condensations would shed light on a number of open questions including: (1) What is the excitation process that produces the strong molecular hydrogen emission? Photoionization, shocks, dissipative MHD waves, and high-energy particles are all possible energy sources. (2) What are the properties of the grains that formed in the cores of the filaments? The Crab serves as a test bed for understanding grain formation in high-redshift galaxies where short-lived massive stars and supernovae must produce the observed dust. (3) What are the molecular processes that occur in this hostile environment? In the ISM $H_2$ commonly forms by grain catalysis, a process dependent on temperature and composition of the grains, but is that true in an environment like the Crab? Many $H_2$ knots do not show dust extinction. Could $H_2$ form without dust? (4) What is the true mass of the filaments? Molecular gas is less emissive than ionized gas and thus the molecular regions of the filaments could contain a substantial amount of mass. (5) The Crab filaments appear to be nearby, well-resolved prototypes of the physically much larger filaments seen in the IGM of cool-core galaxy clusters (Fabian et al. 2008). Both show a low-ionization spectrum, have similar geometries, and are surrounded by ionizing particles. The Crab can serve as a test bed for understanding the physics of the cluster filaments.

In this paper, we explore the physical conditions in the gas that produces the $H_2$ emission. We aim to build on the pioneering first look at this problem by G90. Our approach is to fit photoionization models, supplemented by other sources of heating and ionization, to our extensive observations of a single knot, using the plasma simulation code CLOUDY (Ferland et al. 1998) to explore the $H_2$ excitation mechanisms, formation processes, level populations and the role of dust in the molecular core.

While previous studies have addressed the structure of ionized skins on the filaments (Pequignot & Dennefeld 1983, Sankrit et al. 1998), none have included the molecular cores. In addition, our observations provide us with more detailed constraints than have been used in previous modeling efforts. Many papers have explored a wide range of possible abundances and argued that point-to-point abundance variations within the Crab filament system are quite plausible (e.g. Davidson 1978; Davidson 1979; Pequignot & Dennefeld 1983; Henry and MacAlpine 1982; MacAlpine and Satterfield 2008; Satterfield et al. 2012). Our study uses a wider variety of optical emission lines and focuses on a more spatially compact feature (a knot) than the filaments considered in other papers.

The brightest $H_2$ knot in our catalog, Knot 1, is not the best candidate for this study. It suffers from lying in a confused region of the Crab. It is immediately adjacent to, and probably physically associated with, the much more extensive bright highly-ionized filament FK-10 (Fesen & Kirshner 1982), but FK-10 does *not* have strong $H_2$ emission while Knot 1 is not at all conspicuous in high ionization lines such as [O III] λ5007 (Paper I). It is unclear where the boundaries lie between these two rather different types of emission regions. In addition, the radial velocities determined from our long-slit optical spectra show that there is considerable other material at different points along the same line of sight, and that the strong [S II] emission from Knot 1 comes from material with a very complex velocity field.

We instead study a much simpler region, Knot 51 (K51) from our catalog in Paper II. Figure 1 shows that K51 lies near the tip of a long tendril of filamentary gas. Both K51 and the tendril have small radial velocities, indicating that they lie approximately halfway through the nebula, where the direction of expansion is in the plane of the sky. The tendril is actually a pair of elongated features extending back down into the synchrotron plasma from one of the large-scale filaments that wraps around the outer surface of the synchrotron plasma. The two join into a single tendril before reaching K51. As we discuss below, the velocity structure in this region shows that there is minimal contamination from foreground or background filaments. The situation is still not ideal, because as is described below there is a second $H_2$ knot immediately adjacent to K51 that is part of the same filament, but this is the cleanest case that we can find.

The following sections first present the available observations of K51, and then describe a series of CLOUDY models that explore a large parameter space with the goal of reproducing the observed $H_2$ 2.12 μm emission along with the observed lines of neutral and ionized atomic species. We do not aim to produce an accurate model of the knot, in part because the existing observations do not uniquely specify the geometry of the molecular region (higher spatial resolution data are needed) or reveal the total molecular inventory (sub-mm observations are needed). Rather, this paper is a plausibility study whose goal is to identify what general type of model can account for the existing suite of observations. The observations of ionized and neutral species in K51 serve as a guide to the overall structure of a typical condensation, but in spite of being the simplest case, K51 is still a rather complicated structure for which we do not attempt to fit all details of the observations.

In our previous papers we had assumed that the $H_2$ emission comes from a predominantly molecular core of some type. The simulations presented here show that in the unusual environment of the Crab it is instead possible that the observed $H_2$ emission can be produced by a trace $H_2$ component in what is basically a very extended $H^0$ zone. Although dust is clearly present in the particular case of K51, grains are not actually required to form $H_2$ in this environment, because of the high electron density where atomic hydrogen is present, which results in an enhanced efficiency of the $H^-$ route.

Finally we outline the types of observations that would be needed to further constrain physical models of the atomic or molecular regions.

## 2. Observational Data

We assembled a wide variety of observations of K51 in order to constrain the nature of its ionized, neutral and molecular gas, and its dust content. These include space- and ground-based images at visible, near-infrared (NIR) and mid-infrared wavelengths, together with ground-based visible and NIR spectra.

### 2.1 Narrow-Band Emission Line Images

Figure 2 shows a series of visible and NIR images, all on the same scale, in boxes covering the same 10" (0.097 pc) square patch on the Crab. These are magnified images of the region shown in the insert on Fig. 1. They include K51 as well as Knot 50, which lies 3.4" away. The pair of small white crosses on each image mark the position of the peak $H_2$ emission from the two knots.

The $H_2$ image is in the 2.12 μm line and is from our data described in Papers I and II, taken with the Spartan Infrared Camera on the SOAR Telescope. It has 0.7" FWHM angular resolution. The [S II] λλ6716+6731, [O I] λ6300, [O III] λ5007, Hβ and continuum images are from HST archival data in which the 0.1" pixels undersample the HST point spread function. Since the Crab's expansion causes significant proper motion of the knots, for each filter we have only summed together HST data taken at a single epoch (Table 1). Following the procedure described in Paper II, the world coordinate system of each co-added HST image was then rescaled around the center of the Crab's expansion to adjust it to the same epoch (2009 Dec) as our ground-based $H_2$ images. This allows for an accurate comparison of the Knot's morphology as seen at different wavelengths, smeared only by actual time evolution of its internal structure.

The HST continuum image will be discussed below. The emission-line images in Fig. 2 show that in $H_2$ emission, K51 appears slightly elliptical with a roughly Gaussian surface brightness distribution. The [O III] emission has a conspicuous hole at the position of the $H_2$ centroid and a pair of bright patches on opposite sides of that hole, and then is strong along the filament that runs up to the NW to join onto the main system of high-ionization filaments seen in Fig. 1. The Hβ, [S II] and [O I] images all show the same structure to either side of the molecular core that is seen in the [O III] image, but also have emission from the position of the $H_2$ central core. This follows the general pattern reported by Sankrit et al. (1998) and others, in which low-ionization cores are surrounded by more extended high-ionization halos. In addition, all of the visible-wavelength lines are also emitted by a circular ring to the SE (lower left in Fig. 2) of K51 that includes $H_2$ Knot 50 on the ring's far edge, and additional fainter $H_2$ emission from along the ring.

### 2.2 Spectra

Our grid of long-slit optical spectra taken with the KPNO 2.1 m telescope covers Knots 50 and 51 (Paper II). The spectrum covers the wavelength range λλ3700-7400 Å at 6.3 Å resolution, with a 3.8" slit width and 2.7" FWHM resolution along the slit. Figure 3 (left panel) shows the slit position of the spectrum that crosses Knots 50 and 51, and next to it (in the center two panels) on the same spatial scale are two segments of our KPNO 2D spectrum, showing the various velocity systems in that region of the Crab. Our NIR spectrum across the two knots (from Paper III) is shown in the right-hand box.

The emission lines from Knots 50 and 51 have the same radial velocities, $v_{helio}$ ~ 130 km s$^{-1}$. Regions along the slit close to the two knots also show emission, especially in [O III], from the front and rear sides of the Crab's expanding shell with $v_{helio}$ roughly ±1000 km s$^{-1}$. The blueshifted component of this additional gas falls outside the HST passbands for the emission-line images. The +1000 km s$^{-1}$ component will contribute to the line emission seen in the HST images, but the line-of-sight to K51 is relatively uncontaminated. The spectrum shows that Knots 50 and 51 are at nearly the systemic velocity of the Crab. In addition, where adjacent KPNO slit positions not shown here cross parts of the long double-tendril of emission reaching down to K51 from the NW, the spectra show that the gas all along that tendril has the same radial velocity as K51. From this we deduce that Knots 50 and 51 and the tendril are all part of the same physical structure that is moving mostly in the plane of the sky, and therefore must lie very close to halfway through the overall Crab Nebula.

Knots 50 and 51 are not spatially resolved in our optical spectrum, so we measured them together by extracting a 1D spectrum that covers the 11.2 arcsec region along the slit marked on Figure 3. We also extracted spectra over smaller lengths along the slit trying to isolate Knots 50 and 51, and confirmed that those spectra do not have any major differences between them in their emission-line intensity ratios. Here we have chosen to use the combined spectrum because we know that we have included the total flux from the sum of the two knots.

The measured line strengths relative to Hβ and the Hβ flux are listed in Table 2. The table also lists line strengths corrected for reddening assuming $A_V$ = 1.46 mag (Davidson & Fesen 1985) and a standard Galactic extinction curve. This dereddening is intended to remove the effects of extinction from the ISM along our line of sight to the Crab. The resulting Hα/Hβ ratio is 2.96, very close to the Case B value, implying that there is little additional internal reddening of the Balmer lines. The last column of Table 2 lists the estimated percentage uncertainties for the lines measured from the KPNO spectrum, including the effects of blending, sky subtraction and systematic error in the flux calibration in addition to the pixel-to-pixel signal:noise. Table 2 also lists line strengths relative to Hβ for the total H$_2$ 2.12 μm and Brγ fluxes for Knots 50 and 51 combined together (since the optical line strengths are for the sum of the two knots). These H$_2$ and Brγ line strengths were measured from our NIR images. The H$_2$ line strength is taken from Table 2 of Paper II, and has an estimated uncertainty of ±10% including the random errors and the uncertainties in the sky subtraction and flux calibration. The Brγ line strength was measured for the aperture used for the Hβ spectrum, since the emission occurs over a larger region. The dereddened Brγ/Hβ intensity ratio derived from combining the optical spectra and NIR images is $I(Brγ)/I(Hβ)$ = 0.027, in good agreement with the value 0.028 for Case B. However, it should be noted that the Brγ line is very weak (see figure 3 of Paper II) and so it is not as well measured as H$_2$ or the optical lines.

Finally, we converted the reddening–corrected fluxes to surface brightness. The line emission comes from a much smaller area on the sky than is included in our extracted spectrum (Fig. 3, left panel). The HST images in Fig. 2 show that the Hβ signal:noise ratio is too low to be able to tell exactly where that line comes from, and that the [O III] line comes from a somewhat more extended area than [S II] or [O I]. For purposes of comparison to the simplified models discussed below, we assume that all lines come from the area within our spectroscopic aperture that emits [S II] $\lambda\lambda$6716, 6731, which is 6.4 arcsec$^2$. This is the projected area on the sky, so for comparison to our CLOUDY simulations our measured surface brightness corresponds to a plane-parallel sheet of material seen face on. The relative dereddened line strengths listed in Table 2, scaled to the H$\beta$ surface brightness listed in the footnotes to the table, are the primary constraints on the models that we will describe below. Although the line strengths are derived from the sum of the two knots, the conversion to surface brightness allows us to treat them as if they came just from Knot 51.

We also have a NIR long-slit spectrum across Knots 50 and 51, taken with the OSIRIS spectrometer on the 4 m SOAR Telescope (Paper III). For K51 it shows enough H$_2$ lines for us to measure an H$_2$ excitation temperature $T_{H2}$ = 2837 ± 200K. For comparison, previous studies have shown that the ionized gas has a considerably higher kinetic temperatures, $T_{ion}$ ~ 15,000 K. The 1D-extracted spectrum is shown in Paper III, but here in the last panel of Figure 3 we show the 2D image of the region around the 2.12 μm line. The line has a heliocentric radial velocity $v_{helio}$ = 108±5 km s$^{-1}$, in reasonable agreement with the value $v_{helio}$ ~ 130 km s$^{-1}$ quoted above which was measured from the [S II] lines. This further establishes the spatial connection between the ionized and molecular gas. The 2D NIR spectrum shown in Fig. 3 has 0.7 arcsec spatial resolution, which is high enough to show that Knots 50 and 51 clearly do have the same velocities and in fact may be connected by a bridge of emission. We did not attempt to combine the H$_2$ line fluxes measured from this spectrum with the measurements from our optical spectrum, because the NIR spectrum was taken through a slit only 0.4" wide with a somewhat uncertain placement on the knot.

**2.3 Dust Absorption Features**

The HST intermediate-band continuum image shown in Fig. 2 is of interest because it shows obvious dust absorption (which appears white in our reverse grey-scale image) coincident with K51. We wished to measure the depths of these features in order to find the dust extinction. However, the raw data through the F547M filter include an appreciable contribution from the night sky, which must first be subtracted off. We measured this by using the few existing F547M images that include the outer parts of the Crab together with areas that clearly are outside of the Crab's synchrotron emission region (as determined by comparison to DSS visible and 2MASS NIR images). This let us produce several individual sky-corrected images, but they did not provide high signal:noise coverage of the K51 region, so were not by themselves suitable for this study. However, we were able to use them to calibrate the sky-subtracted surface brightness at a number of points where the synchrotron continuum is not contaminated by emission from filaments. A further problem is that the sky level varied over time. For each image used in our final co-added continuum image, we measured the surface brightness at a series of fiducial points on the nebula that do not include filaments, and worked out flux offsets for each image which then were subtracted to remove relative differences in the sky level. After making those corrections, we combined the images and then finally subtracted a suitable offset from the co-added image so that it has the same surface brightness at each calibration spot as is measured from our individual sky-corrected images. This produced the sky-subtracted continuum image used here. At the position of K51, this sky correction amounts to 44% of the average level of the continuum at points not in the dust features.

The deepest part of the dust absorption feature at K51 has 69% of the flux level of the adjacent (sky-subtracted) continuum regions. We need to account for the fact that K51 is embedded in the synchrotron emitting gas. As noted above, the radial velocity of K51 shows that it must lie approximately half-way through the synchrotron-emitting region, so for our best estimate we subtracted off half of the continuum level to account for emission in our direction originating between us and the knot. After this correction the K51 absorption feature has a central depth of 62%.

There is uncertainty in this measurement due to the patchy distribution of the synchrotron emission. To estimate the effect of this on our extinction measurement, we integrated the continuum brightness along many 2"-wide lines that lie in the plane of the sky and perpendicular to the long axis of the nebula, and formed ratios of the light coming from either side of the midpoint of the nebula. The RMS scatter in these ratios was 5%, which we used as an estimate of the uncertainty in the fraction of the continuum emission that comes from in front of K51, along our line of sight through the knot.

The extended source complicates the conversion from the measured 1.05 mag dip, to the usual $A_V$, which includes scattering as well as absorption. In the approximation that K51 is surrounded by a uniform light source and the optical depth is small, scattering into the line of sight compensates for scattering out of the line of sight, and $A_V$ is larger by a factor of $(1-albedo)^{-1}$. This correction gives us the extinction that is usually compared to the gas column density $N_H$ and which is what would be observed for a sheet of absorbing material in front of a background point source, so we will designate the corrected value as $A_{V-pnt}$. At 5470Å, the albedo is 0.77, 0.67, and 0.81 for the LMC, $R_V = 3.1$, and $R_V = 5.5$ extinction curves given by Draine (2003), corresponding to $A_{V-pnt}$ = 4.6, 3.2, and 5.5 mag, respectively. The actual grain composition and size distribution in the Crab is unknown. In the Cloudy models described below, we have arbitrarily chosen the Orion Nebula grain distribution used by Baldwin et al (1991), for which the albedo at 5470Å is 0.6. This corresponds to $A_{V-pnt}$ = 2.6 mag, which for consistency we use as our observed value for comparison to the models even though it is at the low end of the range of possible $A_{V-pnt}$. This is about the depth where $H_2$ initially forms in PDRs (Tielens & Hollenbach 1985). However, the combination of the unknown grain properties and the uncertain and possibly very large correction needed to the observed absorption depth means that the extinction measurement is only a very marginal constraint.

There is also weaker dust absorption associated with the ring that includes Knot 50. Here the central depths indicate $A_{V-pnt}$ ~ 0.6 −1.3 mag for the same range of assumptions about the albedo of the dust.

**2.4 Mid-IR Images**

There are also archival mid-IR images of the Crab available at 24 μm (Spitzer MIPS), 70 μm (Herschel PACS blue) and 100 μm (Herschel PACS green). These clearly detect the Knot 50+51 complex, but at such low angular resolution that it is unresolved with no measurable structural details. The Knot 50+51 complex is blended with the smeared-out images of surrounding structures, but we have crudely measured the total excess flux from the Knot 50+51 complex above what is our best estimate of the underlying synchrotron continuum. The results for each of these passbands are listed in Table 2. In principal, the wavelength covered by these images should include the thermal emission from the dust in K51; a thermal spectrum peaking at about 80 μm is seen in the integrated light of the Crab (e.g. Marsden et al. 1984) and the models presented below predict dust emission with the correct temperature to produce that emission. Knowing the strength of this feature for K51 together with the $A_{V\text{-}pnt}$ would place interesting constraints on the nature of the dust formed in the Crab (e.g. Sankrit et al. 1998). However, the CLOUDY models described below and the spectra shown by Temim et al. (2012) indicate that the emission from the knots in these bands should be completely dominated by the [O IV] 25.9 μm and [Fe II] 26.0 μm emission lines for the Spitzer MIPS band, and by [O I] 63.2 μm for both Herschel bands.

**2.5 The intensity of the radiation field striking the filaments**

The incident ionizing flux over the 1–5 Ryd energy range is the crucial input parameter for our CLOUDY models. Wu (1981) made careful measurements of the Crab's continuum flux down to 1550 Å, from which it is only a modest extrapolation to 912 Å (1 Ryd). Using Wu's Figure 2, which shows the measured fluxes corrected for extinction E(B-V) = 0.5 ($A_V$ = 1.55 for a standard Galactic reddening curve) and scaled to include the full Crab Nebula, we found log $F_\nu$(912 Å) = -22.71 and $\nu F_\nu$ = 6.5×10$^{-8}$ erg s$^{-1}$. Assuming a distance of 2 kpc, this gives $\nu L_\nu$ = 3.1×10$^{37}$ ergs s$^{-1}$ at 912 Å and an integrated ionizing photon luminosity $Q$(H) = 1.7×10$^{48}$ photons/sec.

We used that $Q$(H) value to calculate the incident flux on a slab of gas that is facing the center of the Crab. For the radial position of K51 (approximately 80% of the way from the center to the edge), and assuming a spherical nebula of radius 1.7 pc with uniform volume emissivity in the synchrotron continuum, this predicts an incident ionizing photon flux on the slab log($\varphi$(H)) = 9.7 (in units of photons s$^{-1}$) on the front face, and log($\varphi$(H)) = 9.0 on the back face. The emission from the photoionized region would be dominated by the front face, so we used the higher value of $\varphi$(H) as a point of comparison with our models.

The actual flux incident on K51 is fairly uncertain. An ambiguity in our incident flux calculation is that the synchrotron emission obviously is *not* uniformly distributed throughout the Crab (as is clear from the HST visible-passband continuum image, for example), plus there could be local beaming effects. The face of the slab will preferentially see the emitting regions directly out in front of it. Also, the incident flux on the slab has a $d^{-3}$ dependence on the distance $d$ to the Crab, which according to Trimble (1973) is probably about 2 kpc but really only is known to lie in the approximate range 1.5-2.3 kpc.

In addition, there are factor-of-three differences between different published luminosity values for the Crab at 1 Ryd. The broad-band SED constructed by Atoyan & Aharonian (1996; and shown in the review article by Hester 2008) smoothly follows the predicted synchrotron SED across the 1-4 Ryd region and corresponds to the lower 1 Ryd luminosity. Atoyan & Aharonian state that this part of their composite spectrum is from Wu (1981). However, the results actually shown in Wu's paper indicate the three times higher value that we have used here. This produces a local bump above the synchrotron spectrum, as is shown in Figure 1 of the review article by Davidson & Fesen (1985).

Wu's result was substantiated by later near-UV observations by Hennessy et al. (1992). We have checked, and the difference in shapes between these two possible SEDs does not affect the model calculations we present below. It is really only the luminosity at 1 Ryd that matters. In summary, we found that a reasonable guess at the ionizing continuum flux is in the neighborhood of $\log(\varphi(H)) = 9.7$ (in units of photons s$^{-1}$ below 912 Å), but with considerable uncertainty.

**2.6 Adopted Error Bars**

This section has described our best estimates of a number of observed parameters that we will next use to constrain simulations of K51, but it is important to keep in mind the rather large uncertainties. Although K51 is the simplest case we could find, it still has a complicated, poorly-known geometry, which limits our interpretation of the measured emission lines strengths. For example, the Hβ line strength is measured to 5-10% accuracy, and so is the $H_2$ 2.12μm line strength, so their intensity ratio in Table 2 is listed as being known to about 10% accuracy. However, these lines come (at least in part) from physically different regions of the K51 complex so it is very hard to know how to compare the measured intensity ratio to the predictions from the models discussed below in which K51 is described as a very simple plane-parallel slab. The same is true for comparisons between emission lines from high-ionization species to those from low-ionization or neutral species; the actual spatial arrangement is far more complicated than that in our simplified models.

The conversion to surface brightness is also uncertain. The deduced surface brightness uses the projected area of the Knot 50+51 complex as if it were a thin sheet seen face-on. If instead the same observed flux in the lines from the ionized gas comes from a thin ionized shell on the surface of a spherical core containing the $H_2$, there would be a factor of 2 times more surface area. The measured emission-line surface brightness also could easily include limb-brightening effects, due to us sighting down a part of the knot's surface that is especially bright because it faces the strongest continuum source. This would affect the lines that are powered by continuum radiation, but it is unclear whether or not the $H_2$ would have this same limb brightening.

In addition, the reddening correction is based in part on measurements of the [S II] intensity ratio made by Miller (1973) at a different point on the Crab. We assume that this corresponds to overlying reddening along the line of sight to the Crab, and have not attempted to correct for internal reddening despite the clear presence of dust within K51. The uncertainties in the geometry also affect our estimates of the incident flux.

All of this emphasizes that we are trying to hit a poorly defined target, which must be kept in mind as we describe our CLOUDY simulations in the following sections. For our comparisons of the simulations to the observations involving surface brightness, we will regard a factor of two agreement to be adequate, in order to account for the geometrical ambiguities. In contrast, the actual observational error bars are appropriate to use for some key emission-line intensity ratios, such as the ratios between different $H_2$ lines that were simultaneously observed through the same spectrograph slit.

**3. CLOUDY Simulations of Knot 51**

We define a successful model of K51 as one that reproduces the observed emission-line surface brightness including the ionized, neutral and molecular species, from within the observed volume of K51. This includes the thermally populated $H_2$ levels discussed in Paper III. Additionally, the line emissivities of [O I] and [S II] should spatially correlate with $H_2$ as is seen in the narrow-band images shown in Paper II. The key issue is to identify the physics that leads to the strong $H_2$ emission observed from the Crab knots.

**3.1 General Methodology**

For our modeling of K51, we used the open-sourced plasma simulation code CLOUDY (Ferland et al. 1998), version 10.00. In one self-consistent calculation, CLOUDY tracks the microphysics and ionization structure through the $H^+$, $H^0$ and $H_2$ zones. This is ideal for modeling these knots within the Crab, for which we have emission-line measurements from all three regions. We assumed that K51 is in steady state, meaning that the timescale for any dynamical changes is much longer than that of any of the atomic processes involved. The implications of this assumption are discussed in Section 4.

The filaments are illuminated by the hard synchrotron radiation field, which we refer to as the incident continuum. For simplicity, we used the synchrotron spectrum averaged over the whole nebula, by digitizing the spectrum shown in Figure 1 of Davidson & Fesen (1985). However we note that the observed SED depends on position, being harder in the central regions and softer in the outer regions (Mori et al. 2004). All models also include the effects of a cosmic ray energy density representing at least the Galactic background level. The "ionizing particles" model discussed below has a far higher cosmic ray energy density representing the particles that produce the Crab's synchrotron continuum emission (see §3.5).

We assumed a plane parallel geometry for all models, because it is the simplest case and the actual geometry is unknown. Since the actual knot is illuminated from all sides, all models are constrained to have a thickness of $10^{16.5}$ cm, which is half the observed width on the sky of K51. The synchrotron continuum penetrates successive $H^+$, $H^0$, and $H_2$ layers. The outer, ionized regions attenuate the incident continuum, eventually allowing $H^0$ and $H_2$ to form. A similar analysis has been conducted by Sankrit et al. (1998), but using a cylindrical geometry, a smaller suite of observational constraints, and, most importantly, not considering molecular gas.

We included the $H_2$ molecule described by Shaw et al. (2005) in our calculations. This model includes all rotational/vibrational levels within the ground electronic state, and the six lowest electronic excited states that are coupled to the ground state. Photoexcitations, collisional excitation and deexcitation, bound-bound transitions, ortho-para conversion, collisional dissociation, high energy effects and the chemistry network are all included, self consistently with other properties of the gas. These calculations also include the model Fe II atom described by Verner et al. (1999).

The Crab contains dust, as is evident from the thermal emission feature peaking at ~80 μm in the SED of the full Crab Nebula (Glaccum et al. 1982; Marsden et al. 1984) and from the dust absorption features seen scattered across the face of the Crab in visible and IR images (Fesen & Blair 1990; Hester et al. 1990). Sankrit et al. (1998) estimated that the Crab filaments might contain up to an order of magnitude greater dust/gas ratio than the ISM, and recent Herschel results (Gomez et al. 2012) appear to confirm this when averaged over a large area on the sky. However, we do not have a strong constraint on the actual dust/gas ratio in the particular knot K51, so we adopted as our default value the ISM dust/gas ratio (although we did check the effect of varying this ratio in our final models; see §3.6).

Dust provides extinction that shields the molecular regions from the incident continuum. Furthermore, dust acts as a catalyst for $H_2$ formation. We adopted the rate given in Cazaux & Tielens (2002) for formation of $H_2$ through grain catalysis. We assumed the same mix of silicate and graphite grains and grain size distribution used by Baldwin et al. (1991) to represent the Orion Nebula dust.

**3.2 $H_2$ Emission in Non-Equilibrium Environments**

One major goal of our models is to reproduce the observed $H_2$ emission strength and thermal level populations presented in Paper III. The $H_2$ lines are very strong relative to nearby H I lines, and the excitation diagrams indicate a temperature of around 3000 K.

$H_2$ emission can be produced by a variety of processes (see Chapter 8 and Appendix 6 in Osterbrock & Ferland 2006; hereafter AGN3). Starlight fluorescent excitation is important in H II regions like the Orion Nebula. The strength of the $H_2$ emission relative to H I recombination lines is determined by the shape of the SED, as discussed in Paper III, with harder SEDs producing weaker $H_2$ / H I ratios. The SED in the Crab is exceptionally hard, but the measured ($H_2$ 2.12 μm)/Brγ intensity ratio is also large. Paper III shows that with the Crab's SED, UV fluorescence photoexcitation and H I recombination will produce a ~3 dex smaller $H_2$ / H I intensity ratio than is seen in the Crab. Fluorescent excitation is negligible in Crab filaments.

X-rays and ionizing particles penetrating into knots' cores are possibilities. These actually behave in a very similar manner (AGN3 Chapter 8). An ionizing particle entering cool atomic or molecular regions will dissociate and ionize the gas. A newly-created free electron has a large amount of energy and creates a shower of secondary suprathermal electrons with a typical energy of about 40 eV. These can excite emission lines, heat the gas, and create further ionization / dissociation. X-rays would produce very similar results. In the central regions of the knots the lower-energy ionizing photons have been removed by the photoelectric opacity of the outer layers, and the penetrating photons have high energy, roughly 10 keV (Figure 4). Inner shells of the heavy elements absorb such high-energy photons. The photoelectron has a large kinetic energy, and the inner-shell hole produces additional high-energy Auger electrons. The result is a shower of secondary electrons producing heating, ionization, and excitation.

The net effect of the secondary electrons produced by X-rays or ionizing particles depends on the degree of ionization of the gas. Collisions between the secondary suprathermal electrons and thermal electrons will change kinetic energy into heat since electron – electron collisions are elastic. In this case the $H_2$ emission will be the result of collisional excitations by thermal electrons and the level populations will appear thermal. When there are few free electrons the energy of the secondary electrons is used to heat, ionize, and excite the gas. Direct excitation of the $H_2$ electronic Lyman-Werner bands will produce non-thermal emission that is nearly identical to that produced by fluorescent excitation. The result in this case will be a highly non-thermal excitation diagram.

Other sources of heat such as shocks or dissipative MHD waves are possible. These generate heat that raises the gas kinetic temperature. The resulting $H_2$ emission will be thermal – if the density is high enough the $H_2$ level populations will have an excitation temperature equal to the gas kinetic temperature. At lower densities the level populations will appear thermal but at a temperature that is smaller than the kinetic temperature.

In the following subsections we describe a series of CLOUDY models that systematically test for these different possible $H_2$ excitation mechanisms.

**3.3 A Basic X-ray Photoionization Model in Pressure Balance**

We first computed a series of CLOUDY models with constant total pressure, photoionized by the Crab's hard synchrotron continuum. The inward pressure of the absorbed synchrotron radiation and the gas pressure at the illuminated face are balanced by the gas pressure within the cloud. The gas is basically in hydrostatic equilibrium (Baldwin et al. 1991). The goal here was to establish a base-line model that would at least properly represent the $H^+$ zone for a simple case.

3.3.1 Setting the Ionization Parameter

First, the ionizing photon flux, $\varphi(H)$, and the initial hydrogen density, $n_H$, were adjusted so that the model reproduces the observed Hβ surface brightness and the density-sensitive [S II] λ6716/ λ6731 ratio (AGN3). The Crab's filaments are made of material expelled from the interior of the progenitor star, and therefore can have very unusual abundances which are known to vary from point to point. As our starting point we adopted the chemical abundances from Model 3 of Pequignot & Dennefeld (1983; their Table 3), which represents gas in the helium-rich band, along with a solar Fe abundance and a solar Ar abundance. For reference, these initial abundances are listed here in Column 2 of Table 3.

For a solar composition H or He absorb most hydrogen-ionizing photons, and there is a simple linear relationship between $\varphi(H)$ and the Hβ surface brightness (AGN3). For highly enhanced abundances, such as those found in the Crab, this is not the case since heavy elements also remove ionizing photons. The $\varphi(H)$ required to reproduce the Hβ surface brightness does depend on the heavy-element composition. As noted previously in §2.5, the observations suggest that log $\varphi(H)$ is roughly 9.7, but with considerable uncertainty. This can be varied to some extent to match observations. This set of abundances gave an optimal value of $\log(\varphi(H)) = 10.06$ and $\log(n_H) = 2.97$ cm$^{-3}$.

3.3.2 Abundances in the Ionized Gas

Determining abundances is not the primary motivation for this work, rather we are trying to understand why there is strong H2 emission. There is an extensive literature about the abundances in the ionized gas, and we do not have useful measurements of the [O III] temperature indicator that would be needed to accurately determine abundances for K51. However, there are known to be point-to-point abundance variations within the Crab (see the introduction), so we did need to carry out a rough determination of the overall metal abundances.

A preliminary adjustment was made to the abundances of O, N, Ne, S, and Ar using the optimizer included within CLOUDY to match to match the strengths of [O II] λ3727, [Ne III] λ3869, [O III] λ4959, [O III] λ5007, [N II] λ6584, [S II] λ6716, [S II] λ6731 and [Ar

III] λ7135 to within a factor of two. The Fe abundance was not allowed to vary because doing so would have greatly slowed the calculations, so we varied the Fe abundance to match [Fe II] λ7155 after using the optimizer. Silicon and chlorine, for which no lines are observed, were included because of their importance to the chemistry network but with solar abundances. The helium abundance was scaled linearly to match the line strengths of He II λ4686 and He I λ5876. The derived He/H is a bit lower than found in some other regions of the Crab such as the helium-rich belt, but within the range of values reported elsewhere. The model reproduces the observed He II / He I ratio. This ratio is mainly sensitive to the shape of the SED (AGN3; Chapter 2), suggesting that the Davidson SED is a reasonable fit to reality. The values of $\varphi(H)$ and $n_H$ were iterated to maintain the observed Hβ surface brightness and [S II] intensity ratio.

These preliminary models still produced [O III], [Ne III] and [Ar III] forbidden lines that were systematically too weak relative to Hβ, so further abundance adjustments were required. Increasing the heavy element abundances will not produce stronger forbidden lines in photoionization equilibrium. These lines are among the strongest gas coolants so raising their abundances will only make the gas cooler, the so-called "thermostat effect." However, *decreasing* all metal abundances raises the gas temperature and shifts the cooling to the optical. This produces stronger optical forbidden lines. Figure 5 shows how surface brightness predictions, along with the $H_2$ temperature and S II ratio, vary as a function of scaling all metal abundances by a given factor. As the metal abundances decrease, the forbidden line strengths increase until the cooling shifts to collisional excitation of Lyman hydrogen lines.

The strengths of these forbidden lines are quite sensitive to the gas temperature. The best abundance set produces $T_e$ = 12500 K in the $O^{++}$ -emitting zone of the ionized gas, very similar to previous modeling results (Davidson 1979, Henry and MacAlpine 1982, Sankrit et al. 1998). Unfortunately, our observations do not detect the [O III] λ4363 line, which could directly measure the temperature. The continuum in this region is both noisy and contaminated by emission from other velocity systems, so only a poor upper limit can be obtained. The models predict that the line should be a factor of two or more below that upper limit.

The best abundance set obtained from this process and used throughout the remainder of this paper is listed in Column 3 of Table 3. It corresponds to scaling all metal abundances down by a factor of 4 relative to those obtained in the preliminary optimization. These abundances lie between the most metal poor (Henry & MacAlpine 1983) and metal rich (Pequignot & Dennefeld 1983) abundance sets found in previous studies.

3.3.3 Sensitivity to the Ionizing Flux

A change in $\varphi(H)$ affects the ionization structure of the knot and also the strengths of higher ionization lines, in particular, [O III] λ5007 and He II λ4686. Larger values of $\varphi(H)$ cause regions that form these lines to extend deeper into the knot. In Figure 6a we show the sensitivity of various quantities with respect to the ionizing flux for the constant pressure model. The ionizing flux cannot increase since it leads to the $H^+$ region becoming substantially larger thereby overestimating [O III] λ5007 and He II λ4686. Decreasing $\varphi(H)$ doesn't provide enough ionizing photons and all lines get weaker, except for the $H_2$ 2.12 μm line which becomes bright.

Despite producing strong $H_2$ emission, this low φ(H) model is *not* able to predict the observed $H_2$ level populations. This is illustrated in Figure 7b, which compares the observed $H_2$ level population diagram to the predicted diagram for a constant-pressure model with log(φ(H)) = 9.66. The observed populations in Knot 51, and also in the other 6 knots observed for Paper III appear thermal, but the higher-energy levels in this series of models have a decidedly non-thermal population. This produces a predicted $H_2$ spectrum that is very different from what is observed. The CLOUDY simulation includes all known resonance and fluorescence processes, and reproduces the relative intensities given by Le Bourlot et al. (1995), Sternberg & Neufeld (1999), and Black & van Dishoeck (1987) (see figure 21 of Shaw et al. (2005)) when the same input parameters are used, so CLOUDY is able to correctly handle the physical conditions found here.

The strong 2.12 μm emission in the low-φ(H) constant pressure model is obtained by combining a very high column density of $H_2$ (nearly $10^{21}$ cm$^{-2}$) with a core that is very dense. The lower temperature is the result of the high density since the cooling per unit volume depends on the square of the density. The observed $H_2$ emission in this case is the result of secondary electron excitation of electronic $H_2$ levels that produces a fluorescence spectrum very similar to starlight pumping. This results in a highly non-thermal excitation diagram, in conflict with the observations, which together with the weakness of the lines from the $H^+$ zone rules out this low φ(H) model.

3.3.4 Final Constant Pressure Model

The constant pressure model that best fits the $H^+$ zone has log(φ(H)) = 10.06 and metal abundances decreased by a factor of 4 from our initial values. We explore it further here. The physical conditions, line emissivities, and hydrogen structure for this model as a function of depth into the cloud are shown in Figure 8a. The top panel shows the number density of the indicated hydrogen state as a fraction of the total hydrogen density including all states. The middle panel provides the electron temperature as well as the electron and hydrogen number densities. The depth dependence of the emissivity, or net emission produced per unit volume at each point and escaping the cloud, is given in the bottom panel of Figure 8a. The surface brightness in each of these lines is not shown in the figure, but is found by integrating the line's emissivity over all depths through the cloud. Examination of Fig. 8a shows that in the constant-pressure model the gas is mostly ionized with very little $H^0$ and no fully molecular zone. The remaining parts of Figure 8 show additional models that will be discussed below.

To quantify the comparison to observations, Table 4 shows the ratio of the predicted/observed surface brightness of various emission lines, and also some key CLOUDY input parameters, for this and the series of additional models described below. The models are compared to the observations by considering predicted/observed surface brightness ratios for each separate line in order to give a clearer picture of which lines are properly reproduced by which models in a situation where the strength of Hβ (typically used as the dominator in tables of emission-line strengths) also varies from model to model. A value of twice the predicted $A_{V-pnt}$ is used in Table 4 because our models only are computed to the center of the knot, while the observational measurement is for the extinction all the way through the knot.

The constant-pressure model matches the observed surface brightness of the higher-ionization forbidden lines and of the recombination lines to within ±30%. However, the [O III], [Ne III] and [Ar III] lines are still systematically 10-20% too weak relative to Hβ, suggesting a residual energy balance problem. There has to be more heating per unit volume that is provided by the observed SED, φ(H), and $n_H$. Previous modeling of the filaments was not so observationally constrained and therefore did not come across this problem. Throughout the process of fitting these models we have accepted cases in which the low-ionization forbidden lines ([O I], [O II], [N II], [S II]) are underpredicted if $H_2$ is also underpredicted, because these forbidden lines are observed to be $H_2$ tracers (Paper II). Our aim is to use the constant pressure model as a baseline to which we will then add an $H_2$ production mechanism which we hope will also contribute lines such as [S II] and [O I]. Indeed, this best-fitting constant pressure model makes a negligible amount of $H_2$ 2.12 μm emission (Table 4), ruling it out as a complete model.

**3.4 Models with Dense Molecular Cores**

The results from the constant-pressure model suggested that a denser core is needed to have enough column to produce the observed $H_2$ emission from within the size of K51 while still matching the lines from the $H^+$ zone. We provided this dense core by introducing a density profile that followed the constant pressure law through the ionized zone, but then quickly ramped up to a higher, constant central core density $n_{core}$ H atoms cm$^{-3}$. The ramp-up started at a depth of $10^{16.37}$ cm which is the point at which most of the [S II] emission has occurred (so that the density determined from the [S II] doublet ratio would not exceed the measured value), and reached the constant value at $n_{core}$ at a depth of $10^{16.50}$ cm.

We experimented with a range of $n_{core}$ values. The results are given in Figure 9, where the solid line shows the predicted/observed $H_2$ intensity ratio. We found that $\log(n_{core}) = 6.0$ reproduces the observed $H_2$ 2.12 μm line. This is our best-fitting model of a dense knot that is powered only by the incident synchrotron continuum from the Crab. The physical conditions, line emissivities, and hydrogen structure for this model are shown in Figure 8b, and predicted/observed line intensities are given in column 3 of Table 4. Note that the line emissivities of Hβ and He I, shown in the bottom panel of Figure 8b, actually increase in the core, where the electron temperature $T_e \sim 500$ K. The predicted line strengths only vary ~20% from Case B showing that these lines are mainly due to recombinations.

This best-fitting dense-core model reproduces the observed H I, He I, He II and higher ionization forbidden lines as well as, and in some cases better, than our best constant pressure model, but now also reproduces the observed $H_2$ 2.12 μm emission. However, like the low φ(H) constant-pressure case discussed above, it conspicuously fails to reproduce the relative strengths of other $H_2$ lines for which measurements or observational upper limits are given in Paper III. This is shown in Figure 10, which displays the predicted/observed ratios for the other $H_2$ lines relative to the 2.12 μm line, as a function of $n_{core}$. The 1-0 S(0) line is predicted to be excited by non-thermal mechanisms and to be several times stronger relative to the 1-0 S(1) 2.12 μm line than is actually observed. This is further illustrated in Figure 7c, which compares the observed $H_2$ level population diagram to the predicted diagram for the $\log(n_{core}) = 6.0$ dense core model. The left most set of points on Figure 10 show that the dense-core case which comes the closest to fitting the observed $H_2$ relative intensities has a lower core density, $\log(n_{core}) = 4.2$. However, Figure 9 shows that this model underpredicts the 2.12 μm line by more than two orders of magnitude so it is not viable.

**3.5 Models with Additional $H_2$ Excitation Mechanisms**

None of the above models get the $H_2$-emitting core hot enough to produce the observed bright, thermal $H_2$ spectrum. Additional heating is needed in the core. As was suggested by G90, the relativistic particles that fill the Crab's synchrotron-emitting plasma are a likely source. We proceeded as in Ferland et al. (2009) and considered two possibilities. The first is a generic heating source corresponding to mechanical energy input, possibly due to shocks or the presence of non-dissipative MHD waves. We will refer to it as the "temperature floor" case. In this model any collisional ionization would be produced by thermal particles so warm temperatures are needed to produce, for instance, $H^+$. The second possibility that we considered is heating by relativistic particles, which we will refer to as the "ionizing particle" case. Ionizing particles, as suggested by their name, can produce ions in gas that has a low kinetic temperature. For more details see the study by Ferland et al. (2009) of the filaments seen in the IGM of cool-core galaxy clusters.

3.5.1 Temperature floor case

In the temperature-floor case, the electron temperature is artificially fixed at the observed $H_2$ temperature, $T_{H2}$ =2800 K, corresponding to a heating rate of $2.5 \times 10^{-16}$ ergs s$^{-1}$ cm$^{-3}$. The only source of ionization in the deeper zones of the knot is through thermal collisions, and the dissociation of $H_2$ occurs only when the temperature is quite high. We included ionization by cosmic rays at the Galactic background rate of $2 \times 10^{-16}$ s$^{-1}$ (Indriolo et al. 2007), but not the effects of the additional high-energy particles localized in the Crab Nebula. We iterated over a range of core densities until we reproduced the observed $H_2$ 2.12 μm surface brightness. From this procedure, we found the optimal core density to be $n_{core} = 10^{5.1}$ cm$^{-3}$.

3.5.2 Ionizing particle energy density

In the ionizing particles case, we included the effects of the high-energy particles that *must* pervade the Crab's synchrotron plasma. We assumed that these particles penetrate into the dense knots, heating, ionizing and exciting the gas. In regions with high electron fractions, energetic particles predominately heat the gas, while in regions with low electron fractions, they predominately ionize and excite the gas (AGN3; Ferland et al. 2009). We add ionizing particles as in Ferland et al. (2009). CLOUDY uses the results of Dalgarno, Yan & Liu (1999) to include the effects of cosmic rays. The key input parameter for CLOUDY in this process is the ionization rate $\xi_H$ per $H_2$. We parameterize this rate as a ratio relative to the ionization rate in the ISM, $\xi_0 = 2 \times 10^{-16}$ s$^{-1}$.

We ran CLOUDY models for a large grid of ionization rates per particle $\xi_H$ and core densities $n_{core}$ to find the models which reproduce the observed $H_2$ 2.12 μm surface brightness and the observed $H_2$ temperature. Figure 11 compares the constraints placed by the observed $H_2$ temperature and by the Hβ and $H_2$ surface brightnesses in the $\xi_H / \xi_0 - n_{core}$ parameter space. The slanting dashed line marks models which reproduce the observed temperature $T_{H2} = 2800$ K, with the gray shaded area indicating the range due to the observational uncertainty in the temperature (as given in Paper III). The solid contours show the logarithm of the predicted/observed $H_2$ surface brightness, while the dotted contours show the logarithm of the predicted/observed $H\beta$ surface brightness. A perfect model would have the dashed line simultaneously intersecting with the 0.0 contours for both $H_2$ and $H\beta$, but in fact all models overpredict $H\beta$. We adopted a scale factor of $\xi_H / \xi_0 = 10^{5.3}$ and a core density of $n_{core} = 10^{5.25}$ cm$^{-3}$ to be the best parameters for the ionizing particles case. This model overpredicts the $H\beta$ surface brightness by approximately a factor of 2.

We next check to see if the deduced value $\xi_H / \xi_0 = 10^{5.3}$ is reasonable. Positively charged nuclei may be an important component of the ionizing particles (e.g. Hoshino et al. 1992; Gallant & Arons 1994), but there is no way to directly measure their presence. For the electrons and positrons, on the other hand, the energy distribution can be determined from the observed synchrotron radiation, so in the following discussion we will consider just the contribution of these particles. In the case of the ISM this underestimates the cosmic ray energy density by about 1 dex (Webber 1998).

The number of ionizations that the electrons and positrons will cause within the knot depends on the degree to which they are trapped by tangled magnetic fields. One limit is to assume that the magnetic field threads though the knot simply so that particles make only one pass through the knot. Then the flux of particles inside the knot is the same as the flux on the outside where it can be deduced from the synchrotron radiation. Using the number distribution of electrons as a function of energy found by Atoyan & Aharonian (1996) and the ionization cross section of atomic H (Indriolo, Fields, & McCall, 2009) scaled up by a factor 2.3 to account for ionization by secondary electrons and the difference between H and $H_2$ (Glassgold & Langer 1974), and integrating over all energies above 50 keV, the minimum energy for which a particle reaches the center of the knot, we found $\xi_H / \xi_0 \geq 10^{3.2}$. This is 100 times smaller than is inferred from the CLOUDY models.

An upper limit from the electron/positron contribution is to assume complete trapping so that all of the energy carried by ionizing particles entering the face of the knot is used up in heating or ionizing the large column density of neutral gas. Assuming energy equipartition with a 300 μG magnetic field (Marsden et al 1984)[6] the ionizing particle energy density in the synchrotron plasma is $U_{IP} = 2000$ eV cm$^{-3}$. The energy flux into the knot is $cU_{IP}$, and assuming that there is about one ionization per 40 eV deposited, the total ionization rate for a knot of half-thickness $l_{knot} = 10^{16.5}$ cm will be $\xi_H \sim (c\, U_{IP})/(40\, l_{knot}\, n_{core}) \sim 1 \times 10^{-9}$ s$^{-1}$. This corresponds to $\xi_H / \xi_0 \leq 10^{6.8}$, about 30 times larger than the model requirements. The value $\xi_H / \xi_0 = 10^{5.3}$ found for our ionizing particle model lies within the range set by these two limiting cases.

For energy equipartition with the magnetic field, the energetic electron energy density corresponds to a pressure of $8 \times 10^6$ K cm$^{-3}$. This is almost two orders of magnitude less than the gas pressure in the core, so even for the case of maximum trapping the ionizing particles are not dynamically important and hence do not affect the geometry of the gas.

---

[6] Recent estimates for the magnetic field at the radial distance of Knot 51 range from 240 μG to produce the synchrotron emission (Atoyan & Aharonian 1996) to 540 μG in the MHD simulations by Hester et al. (2006).

### 3.5.3 Extended $H^0$ zones

A crucial result for both cases is that hydrogen in the core of K51 is almost entirely atomic; only a trace amount of $H_2$ is required to reproduce the observed 2.12 μm line. The lower panels of Figures 8a-8d show the emissivity as a function of depth for several key lines. In the temperature floor and ionizing particle cases, the Hβ and [S II] line emissivities increase in the $H^0/H_2$ zone. The middle panels in Figure 8a-8d displays the temperature and density as a function of depth.

We also found that less than half of the $H_2$ formation proceeds on grains. $H_2$ is mainly formed through associative detachment: $H^- + H^0 \rightarrow H_2 + e^-$ (see Ferland, Fabian & Johnstone 1994). The main $H_2$ destruction paths include UV fluorescence and reverse molecular reactions involved in the formation of $H_2$. The dependence of these formation and destruction mechanisms on depth is shown in the middle panels and bottom panels, respectively, of Figure 12.

### 3.5.4 Heating and cooling

Figure 13 identifies contributors to the total heating and cooling for both cases. Photoionization of H or He is the dominant heating process in the knot until a depth of $2.7 \times 10^{16}$ cm where the temperature floor is imposed in the temperature floor case, and to a depth of $6.3 \times 10^{15}$ cm in the ionizing particles case. The heating sources in the core for the temperature floor model are not meaningful, because the temperature has been artificially set to a fixed value. In the ionizing particles case, the $H^0/H_2$ zone is heated primarily by the ionizing particles, followed by a 10% contribution by ionization of $Fe^0$. Cooling proceeds mainly though electron collisional excitation of [O III] lines in the $H^+$ zone and of [Fe II] lines in the $H^0/H_2$ zone for the ionizing particles case.

### 3.5.5 Effect of the ionizing flux in extra heating models

As explained in §3.3.3, the ionizing photon flux, φ(H), most greatly affects higher ionization lines, in particular, [O III] λ5007 and He II λ4686. Larger values of φ(H) allow regions that form these lines to extend deeper into the knot. Figure 6 shows the dependence of various lines and physically relevant parameters on φ(H) for the temperature floor and ionizing particle cases. We used log(φ(H)) = 10.06 in the models presented in Table 4. We note that Hβ and He I are not strongly correlated with φ(H) in the ionizing particles case since cosmic-ray heating accounts for over 50% of the surface brightness. Similarly, the gas kinetic temperature weighted by the $H_2$ abundance at each point, $T_{H2}$, is almost entirely set by ionizing particle heating with the ionizing photon flux having only a small effect.

### 3.5.6 Results for models with additional $H_2$ excitation

Table 4 shows that the core temperatures, $H_2$ column densities and $A_{V\text{-}pnt}$ for the temperature floor and ionizing particle models are very similar. Both models produce a satisfactory $H_2$ emission spectrum (as shown in panels (d) and (e) of Figure 7). The models are largely equivalent: the ionizing particles have provided a physical reason for the temperature floor.

The differences in the results from these two models come from their ability to reproduce the observed neutral and recombination line emission. In the ionizing particle case, the [O

I] emission is correct to within 30% and [N I] is overpredicted by 60%. Ionizing particles ionize and heat the gas, producing stronger lines in the neutral zone. In this regard, the ionizing particles case is more successful than the temperature floor case which produces only about 20% of the [O I] and [N I] emission. The reason for this stems from the temperature at which the $H_2$ is observed. Neutral oxygen requires temperatures of about 7000-8000 K to thermally excite substantial emission, while the measured $H_2$ temperature is only about 2800 K. In the temperature floor case, the [S II] doublet is underpredicted by a factor of 3, however in the ionizing particles it is correct to within ~15%. The extra [S II] emission in the ionizing particles case comes from the core, which we explain in detail below. Finally, the temperature floor case correctly reproduces the [O II] λ7320 line but the ionizing particles case overpredicts it by a factor of two. However, there is considerable uncertainty in this particular measurement.

Conversely, the $H^0$ and $He^0$ recombination lines are overpredicted by approximately a factor of two in the ionizing particles case. This is a failure of the ionizing particles case, but the temperature floor case correctly predicts all recombination lines to within 20%. The discrepancy here arises from the physics in the core. Figure 8 displays the emissivity of Hβ for both cases (shown as the dashed black line in the bottom panels). The ionizing particles case features an increase in Hβ emission in the core due to ionization of $H^0$, the majority of the hydrogen, followed by recombinations. The temperature-floor case does not display this feature since the gas is too cool to collisionally ionize H. In this aspect, the temperature floor case is more successful than the ionizing particles case. The existing HST Hβ images have too low signal/noise ratio to be able to tell whether or not there is emission from the core. Future high-resolution Hβ imaging with better signal/noise ratio will help to decide whether the ionizing particle case is a viable model.

A successful feature of both the temperature-floor case and the ionizing-particles case is that part of the [S II] λλ6720 doublet is emitted from the same region as molecular hydrogen (bottom panel of Figure 8c and 8d). This is observed to be the case in Knot 51 (Fig 2), and in Paper II we showed that many other, fainter regions of the Crab's system of filaments produce $H_2$ emission that correlates with [S II]. Our two viable models are consistent with this observation. Conversely, Figure 8a shows that the constant pressure model does not emit any $H_2$ and Figure 8b shows that the dense core model emits $H_2$ only from a region largely separate from that which emits [S II].

Both the temperature floor and ionizing particle cases are able to reproduce the major observational signatures of the $H_2$ knots. However, each of these models has failures as well as successes, making it unclear which model is to be preferred. In both cases, we are able to successfully reproduce the forbidden lines from the ionized gas to within a factor of two, the $H_2$ 2.12 μm line to within 10%, and the correct relative strengths of the observed $H_2$ lines.

A variation on the ionizing particles model is to accept a worse fit to the $H_2$ 2.12 μm surface brightness, but one that is still within the uncertainty of the measurement, in order to lower the recombination line strengths enough so that they fit the observations to within a factor of two. A model with $\log(\xi_H / \xi_0) = 10^{5.2}$ and $\log(n_{core}) = 5.1$ produces only 70% of the observed $H_2$ emission, but fits all the optical lines to within a factor two except for [O II] λ7320. However, the purpose of this paper is to explore models that match the H$_2$ emission. Therefore, this model which purposefully did not attempt to fit the H$_2$ emission will not be considered any further.

### 3.6 Models with Increased Dust Abundance

The models with extra H$_2$ excitation mechanisms reproduce the observed H$_2$ spectrum along with the surface brightness of many optical lines. However, these models underpredict the observed $A_{V\text{-}pnt}$ ~ 3 mag. Recent estimates of the Crab's total dust content range from 0.002 M_sun (Temim et al. 2012) to 0.25 M_sun Gomez et al. 2012), and the dust/gas ratio is known to vary from place to place within the filamentary structure (Sankrit et al. 1998). Sankrit et al. (1998) estimate the mass ratio $M_{dust}$ / $M_{gas}$ to be as large as ten times the ISM value for one filament. We do not have additional information that allows us to constrain the dust abundance in the Crab. However, we tested the robustness of our results by varying the dust abundance, ranging from dust-free to an order of magnitude above the ISM value. For simplicity, in the ionizing particles case, we did not reoptimize the heating rate and core density as this is a computationally expensive task. Instead, we simply explore the parameter space here to present the sensitivity of various relevant values.

The top panel of Figure 14 shows the fraction of H$_2$ formation due to grain catalysis ("grains") and associative detachment ("H$^-$ + H$^0$") at the center of the knot and the predicted/observed $A_{V\text{-}pnt}$ ratio measured clear through the knot, plotted as a function of dust abundance for the temperature floor and ionizing particle cases. In both cases, the H$_2$ formation rate by grain catalysis eventually becomes faster than associative detachment. The observed $A_{V\text{-}pnt}$ is found at approximately 10 times the ISM dust abundance. The bottom panel of Figure 14 shows a slow decrease in the optical line strengths as function of dust abundance due to increased extinction. In contrast, H$_2$ emission increases with dust abundance due to an increase in the H$_2$ abundance due to grain catalysis and shielding.

An ionizing particles model with a dust abundance a factor of 5 above the ISM value gives a reasonable fit to observed surface brightnesses as well as matching the observed $A_{V\text{-}pnt}$ to within about a factor of three (Table 4). Internal extinction removes many of the Hβ photons generated in the central core and brings this line into better agreement with the observations. This model matches all the higher ionization forbidden lines to within 20%, all recombination lines to within at least a factor of 2, and the thermal H$_2$ spectrum (Figure 7f). However, it does fail to produce enough [N II] λ6584 and [O II] λ7320.

Table 4 does not include the Spitzer or Herschel broad-band measurements because these are extremely crude measurements which combine continuum emission with a variety of emission lines, and therefore do not deserve the same weight as the individual lines listed in the table. However, the predictions from the extra heating models are 2–5 times too low for the temperature floor model, within a factor of two for the ionizing particles model, and 1–4 times too high for the 5× dust model. We regard all of these as being satisfactory matches.

We briefly explored a model with no dust at all, to determine whether sufficient H$_2$ would be formed by routes not involving dust. Although it is clear that K51 must contain some amount of dust because we see dust absorption in the HST continuum image (§2.3), there are other H$_2$ knots in the Crab that do not have detectable dust absorption and which therefore might be described by such a model. As Figure 14 (bottom panels) shows, models without dust can still produce strong $H_2$ emission. For example, a dust-free model with density $n_{core} = 10^{5.2}$ cm$^{-3}$ at the center of the knot and a temperature floor is able to correctly reproduce the observed $H_2$ 2.12 μm surface brightness and thermal $H_2$ spectrum. It underpredicts [N I] λ5198, [N II] λ6584, [O I] λ6300, and [S II] λ6720 by about the same amount as in the temperature floor model described above (in which dust is present at the ISM abundance). This dust-free model is *not* intended to describe K51, but it does show that in principle dust-free knots could form $H_2$ and produce strong $H_2$ emission.

## 4. Discussion

### 4.1 Understanding the $H_2$ Forming Environment

For both the temperature floor and the ionizing particle models, we find the surprising result that the strong $H_2$ emission comes from a region where hydrogen is predominantly atomic. The overwhelming reason is that the core is insufficiently shielded from the Balmer continuum to allow $H_2$ to fully form. The energetic photons penetrate into the core and produce a region where hydrogen is mainly atomic but there is a significant electron density.

The $H_2$ formation mechanism in this environment is primarily associative detachment (H$^-$ + H → $H_2$ + e). This is because the product of the electron density $n_e$ and the hydrogen density $n_H$ is large enough to allow $H_2$ formation to proceed faster by associative detachment rather than by the usual grain catalysis method (2H + grain → $H_2$ + grain). For typical conditions present in the ISM at the H$^0$/$H_2$ interface,

$$\frac{r_H}{r_{grain}} = \frac{n_e \alpha_H}{n_H \alpha_{grain}} \approx \frac{n_e}{n_H} 250 \qquad (1)$$

(Ferland et al. 1994 equation A10), where $r_{grain}$ and $r_H$ are the rates, and $\alpha_{grain}$ and $\alpha_H$ are the rate coefficients for grain catalysis and associative detachment, respectively. For $n_e/n_H > 4 \times 10^{-3}$ associative detachment is faster than grain catalysis at forming $H_2$. In the cases we have considered in this paper, $n_e/n_H \sim 10^{-2}$ in the core of the knot, showing that associative detachment is dominant. This result will hold regardless of any changes to the abundances or φ(H). If we are overpredicting the $H_2$ emission due to uncertainties in the geometry, then the core density $n_{core}$ would need to be lowered to compensate, producing a larger $n_e/n_H$ and still favoring associative detachment.

### 4.2 Timescale Considerations

We have assumed that the age of the SNR is long enough for physical processes occurring in K51 to be in steady state. Here we discuss processes for which this is not true: (1) the known variability in the incident continuum spectrum due to high-energy gamma ray bursts; (2) the $H_2$ formation timescale; (3) the dynamical timescale; (4) the evaporation timescale.

The gamma-ray activity of the Crab is the quintessential example for high-energy plasma physics, and as such, has been the subject of several recent investigations (e.g. Abdo et al. 2011; Tavani et al 2011). In order to consider the most extreme case in which a gamma-ray flare would alter the physical conditions in which $H_2$ forms, we have chosen an exceptionally energetic flare to compare with the total luminosity produced by the pulsar. Buehler et al. (2012) detected a gamma-ray burst that corresponds to a peak luminosity of ~$4 \times 10^{36}$ erg s$^{-1}$. However, that gamma-ray peak only corresponds to ~5% of the total luminosity currently being injected into the nebula, after the luminosity responsible for acceleration of the filaments is removed. Such slight variations are negligible when considering the effects on the molecular knots.

Of greater concern is the H$_2$ formation timescale. In K51, this is set by the associative detachment reaction. The rate for this reaction is limited by the rate for radiative attachment, H$^0$ + $e^-$ → H$^-$ + $\gamma$. Therefore, from Equation 1 the rate is given by $r_H = n_e \alpha_H$ where $\alpha_H = 8.861 \times 10^{-18} T_e^{0.663}$ is the rate coefficient for radiative attachment at the temperature in core of the Crab (Ferland et al. 1994). The electron (or kinetic) temperature in the core is very close to the H$_2$ temperature of 2800 K. The smallest electron density in the core is ~$7 \times 10^2$ cm$^{-3}$ in the temperature floor case, and this will set the most extreme timescale. In this situation, $\tau_H = 1/r_H$ ~ $2.5 \times 10^4$ yr, which is much longer than the age of the Crab Nebula. This indicates that future calculations must include time-dependent chemistry. We hope to explore the effects of a time-dependent calculation in a future paper.

Our extra-heating models are highly over-pressured in the core relative to the envelope of ionized gas, raising the question of how long such knots could survive. The sound crossing time is set by the sound speed in the core, $c_s = (kT/\mu_0 m_H)^{1/2}$ ~ $7.0 \times 10^5$ cm s$^{-1}$, and is $t_s$ ~ 1400 yr, roughly the age of the Crab. However, the evaporation (or mass-loss) timescale is more physically relevant. Assuming that the heated gas in the ionized layer flows outward at the speed of sound in the ionized gas, $c_{s\text{-}ion}$ ~ $1.4 \times 10^6$ cm s$^{-1}$, the mass loss rate is $dm/dt$ ~ $4\pi l_{knot}^2 c_{s\text{-}ion} \rho_{ion}$ where $\rho_{ion}$ is the total mass density in the ionized skin. Using a total K51 mass of $2 \times 10^{-3}$ M_sun (as found below in §4.6) gives $t_{pe}$ ~ $m/(dm/dt)$ ~ 1800 yr. Therefore, even if K51 formed instantaneously after the supernova explosion, it is reasonable that it has marginally survived.

A related possibility is a dynamical environment where dense seeds are ablated to produce the emission in a non-equilibrium flow. This is highly similar to the knots in the Helix planetary nebula (Henney et al. 2007).

**4.3 Effects of Geometrical Uncertainty**

We assumed a plane-parallel geometry for our modeling of K51, and compared our results to emergent line intensities deduced as if we were in fact measuring a plane-parallel slab face-on. In reality, although K51 is the simplest case available, it has a complicated and poorly defined morphology. Our modeling does not intend to accurately describe all of the features present in K51. Instead, we offer a proof of concept model in which our main results hold regardless of the actual geometry. As an example of the uncertainties due to the poorly-known geometry, we consider the effects if the actual geometry were spherical as opposed to the assumed plane-parallel situation.

As described in §2.2, we determined an observed surface brightness for each emission line by taking the total flux measured from the knot and dividing by the knot's projected surface area on the sky. But a sphere of a given radius has two times more surface area than the front and back surfaces of a thin disk of the same radius, so the emergent energy per unit area would be two times smaller.

In our plane-parallel models, a simulation stopping halfway through the knot has an $H^0/H_2$ core that extends over a range of depth that is one-third the depth of the ionized zone, producing a core:ionized volume ratio of 1:3. But if this same dependence on depth into the cloud were true for a spherical cloud, the core:ionized volume ratio would become about 0.02. Thus, if our CLOUDY results were adjusted to account for a spherical geometry, the predicted ratio of $H_2$:ionized surface brightness should be decreased by approximately a factor of 16.

An additional uncertainty comes from the ~1 mag of internal extinction at visible wavelengths. If the knot is opaque to optical lines then emission lines coming from the rear half of the knot (which we assume to be illuminated from the rear side) would need a greater reddening correction. However, the knot is transparent to IR lines, thus, only half of the volume emitting optical lines from the ionized outer skin would be observed, while all of the $H_2$-emitting volume would be seen at NIR wavelengths. This would improve the fit for the ionizing particles case, since the overpredicted $H^0$ and $He^0$ visible-wavelength lines originate mostly from the core, and on average would experience greater extinction than has been accounted for and hence would become weaker. Although in that case the $Br\gamma/H\beta$ ratio would no longer fit the Case B values, the measurement of the $Br\gamma$ line strength has considerable uncertainty and is not a tight constraint on our models.

These factors may or may not partially cancel each other out, but taking them into account would change our optimization of the abundances, of $\varphi(H)$, and of the density law. However, we stress that such a reoptimization is unlikely to change our basic results about $H_2$ formation and the state of hydrogen in the core.

**4.4 Are models with fully molecular cores ruled out?**

In Papers I and III we presented estimates of the $H_2$ emissivity per H baryon based on the assumption that the emitting cores are fully molecular and are excited solely by collisions. We suggested that our $H_2$ survey selectively found molecular regions in which the temperature and density are in the range where $H_2$ has its highest emissivity. To create a fully molecular core, the internal extinction must be large to shield UV radiation from inducing dissociation (Tielens & Hollenbach 1985). As is discussed in Paper III, given the observed 2800 K $H_2$ temperature, the measured $H_2$ emission would need to come from a much smaller volume than is contained within the ~1" diameter structures seen in our $H_2$ images, suggesting that the emission might come from either a thin shell around a completely obscured inner core or from dense sub-condensations with a very small volume filling factor.

Here, we have arrived at a very different picture. At least in the case of K51, the Crab SED is insufficiently shielded to neglect photodissociation. The two models which include additional excitation mechanisms produce an extended $H^0$ zone in which hydrogen is almost entirely atomic, not molecular, but has a sufficiently large column density so that the observed $H_2$ emission still can be produced within the observed volume of the knot. The measured $H_2$ emission comes from an atomic region that fills the volume of the knot, rather than from the fully molecular zone posited in Paper III.

However, the statistics suggest that other $H_2$ knots may well have fully molecular regions. In Paper II we found that the $H_2$ knots are predominantly redshifted (34 have $v > 100$ km

s$^{-1}$ while only 7 have $v < -100$ km s$^{-1}$). This means that the knots we have detected in the H$_2$ line are on the back side of the expanding nebula, which has a systemic velocity near 0 km s$^{-1}$. Our translucent model does not account for this observation. A possible explanation is that in most H$_2$ knots the excitation is stronger on the side facing the pulsar and that a fully molecular zone with significant 2.12 µm extinction has formed on the side away from the pulsar, which for the missing blueshifted knots would be the side nearer to us.

In addition, some of the other knots found in our H$_2$ survey could turn out to be fully molecular gas that is optimally emitting in H$_2$ as suggested in Paper III. The extended very faint H$_2$ emission that seems to follow many of the large-scale filament structures (Paper II) could represent fully molecular regions with less than the maximum H$_2$ emissivity. There could also be additional discrete condensations of cooler molecular gas that do not produce detectable H$_2$ emission.

An important check on the possible existence of fully molecular gas will come from future mm-wavelength observations. Figure 15 compares the very different appearance of the mid-IR spectra expected from an extended H$^0$ zone of the type we propose for K51, and from a fully molecular core. For both cases we show the predicted spectrum from an H$_2$ knot, including the line emission, bound-free emission, free-free emission and dust continuum emission produced within the knot, together with the transmitted part of the incident synchrotron continuum. The mm-wavelength range in the fully molecular model (lower panel) has a spectrum rich with lines from CO, CS, HCN, and a host of other molecules, which are mostly absent in the temperature floor model (upper panel). As an example of the difference between the models, for the temperature floor and ionizing particle models the predicted strength of the CO j=1-0 line at 2.6 mm is ~10$^7$ times fainter than H$_2$ 2.12 µm. This corresponds to a flux of about 0.6 mJy km s$^{-1}$. For a line width of 20 km s$^{-1}$ the peak brightness would be 0.05-0.24 mJy, which is very faint but probably just detectable with ALMA if the line is narrow and all of the flux falls in one beam. In the fully molecular core case, the CO emission would instead come from a region where H$_2$ is not emissive because it is much cooler than the 2800 K that we found for K51, so for the same amount of H$_2$ emission the CO lines would be extremely bright.

In Paper III we suggested a model in which the H$_2$ emission comes from a thin, heated outer layer on a cool core of molecular gas. In that sort of picture, the column density through K51 of warmer (emitting) H$_2$ is 2x10$^{18}$ cm$^{-2}$, and (based on the ratio of the knot diameter to the thickness of the H$_2$-emitting layer) the total column density including the cold (non-emitting) H$_2$ is roughly 2x10$^{21}$ cm$^{-2}$. Using a conversion factor $X = 3x10^{20}$ H$_2$ cm$^{-2}$ (K km s$^{-1}$)$^{-1}$ (Young & Scoville 1991), the surface brightness in the CO line should be about 7 K km s$^{-1}$. This is 10$^4$ times brighter than the previous case, so CO would be relatively easy to detect with current interferometers. However, this latter result assumes that the CO-emitting gas fills the entire volume corresponding to the observed 2" diameter of K51; a smaller filling factor (as is likely to be the case) would make the line much harder to detect. Also, the *X*-factor value is based on measurements of optically thick gas with typical ISM metallicities, so it is unclear how accurate it is for the case of the Crab. In spite of these two caveats, a deep CO measurement is likely to provide a critical test of our model.

## 4.5 The Role of Dust

Thermal emission from dust is clearly seen in the integrated continuum spectrum from the full Crab Nebula (Marsden et al. 1984). Temim et al. (2006, 2012) estimated the total dust mass associated with the filament system to be $10^{-3}$–$10^{-2}$ M_sun while Gomez et al. (2012) estimated the range to be 0.26-0.68 M_sun. Well-resolved dust absorption features are known to be present at a number of points across the Crab (Fesen & Blair 1990, Hester et al. 1990). Previous photoionization models of the filaments have yielded a dust to gas ratio estimate an order of magnitude above the ISM (Sankrit et al. 1998).

Our extra heating models correctly predict the shape of the observed thermal emission bump from dust in the Crab. The filled circles in the upper panel of Figure 15 (plotted on top of the temperature floor model) show observed points derived from Herschel, Spitzer, Wise and Planck images, taken from Table A1 of Gomez et al. (2012). The Gomez et al. (2012) results are integrated over an aperture that covers most of the Crab. They are corrected for the average line emission, but include a very strong contribution from the unfiltered synchrotron emission. We adjusted their observed flux values by subtracting off the synchrotron component, scaling the remaining dust emission to match the height of our predicted dust emission, and finally adding back on the underlying continuum component predicted by our models. This means that the observed points at $\lambda < 10$ μm and $\lambda > 500$ μm have been forced to match our simulated spectrum, but that the points between those two wavelength limits show that the shape and peak wavelength of our predicted dust emission closely match the observations. This tests the dust temperature that is predicted by our models, but not the dust abundance. Our models predict that as a result of the assumed grain-size distribution and compositions, the dust will have a range of temperatures lying between 38 and 54 K. This is in reasonable agreement with the cold and hot components found by Gomez et al (2012), which had $T_{dust} \sim 28$ and 63 K, respectively.

While we do not have observations that allow us to directly constrain the dust abundance in K51, we explored the effects of varying this quantity in §3.6. The best model from this analysis arises from adopting a dust abundance five times greater than the ISM. This creates an environment in which $H_2$ forms at equal rates via grain catalysis and associative detachment, emphasizing the importance of less common $H_2$ formation routes even in dustier environments.

As was noted in §2.3, we assumed dust with optical properties that produce a rather low albedo. If we were to repeat the entire analysis using dust properties at the other end of the albedo range, we would correct the observed absorption dip to a significantly larger $A_{V-pnt}$ value but then the models would also predict a larger $A_{V-pnt}$. The two $A_{V-pnt}$ values should match about as well as they do for the dust properties assumed here, but there would be changes in the dust's effect on the internal structure of the knot and the emission line spectrum which we would expect to be fairly modest.

Are there dust-free knots in the Crab? Although K51 clearly contains dust, archival HST continuum images show that there are other $H_2$ knots that do *not* appear to be associated with specific dust absorption features. Therefore, it is possible that the observed dust emission originates only in some parts of the filamentary system, while other parts have no dust. The dust-free model briefly described in §3.6 represents a first pass at describing dense condensations in such regions. It is able to produce the observed 2.12 μm line in a unique environment in which grain catalysis, typically a primary mechanism for $H_2$ formation in the ISM, is absent and associative detachment is the only major $H_2$ formation mechanism. Our dust-free model is roughly as successful as the temperature floor model in describing the observed emission line intensities from K51.

**4.6 Mass Estimate**

Previous sections have developed a type of model that largely reproduces the observed emission from the ionized and molecular regions of K51. Here we use this model to estimate the total mass in the knot. Previous studies have used hydrogen recombination lines to estimate the mass of ionized gas, and $H_2$ lines to measure the mass in molecular regions. The latter is especially uncertain because we do not have any direct measurement of the density of the molecular core, and the derived mass depends on the density when the density is low.

We use our model to convert line intensities into a mass. In this approach, the model self consistently accounts for regions where hydrogen is ionized, molecular, *and atomic*. The latter can only be estimated with a model because there are no existing 21 cm detections of the knots[7]. The model reports the total column density required to produce the surface brightness in H I or $H_2$ lines, which we then rescale to subtract off the $H^+$ zone. A surface brightness can then be converted directly into a column density and, if the size of the knot is known, into the total mass, for the atomic and molecular regions.

We can use the temperature floor and/or ionizing particle model to estimate the total mass that is needed to produce the observed $H_2$ emission. To within 5%, the two models give the same result. Only a modest amount of $H_2$ is required, because the density and kinetic temperature in the extended $H^0/H_2$ core are in the range over which $H_2$ will emit with nearly maximum efficiency (see Figure 5 of Paper III). However, both models predict that there will be a factor of 2000 times more H atoms present in the form of $H^0$ than in the form of $H_2$, which is a significant amount of mass that would not otherwise be detectable.

For K51 alone, using the ionizing particle model, the mass of $H_2$ is $m_{H2} = 8 \times 10^{-7}$ M_sun while the total mass in the core including $H^0$ and the He and heavy elements in the $H^0/H_2$ zone is $m_{core} = 2 \times 10^{-3}$ M_sun. For the sum of all 55 $H_2$ knots reported in Paper II, scaling by the observed $H_2$ flux gives $m_{H2,total} = 4 \times 10^{-5}$ M_sun and $m_{core,total} = 0.1$ M_sun. This assumes that the surface brightness to mass conversion factor derived for K51 applies to all knots. The latter value is an order of magnitude smaller than the estimate made by Fesen, Shull & Hurford (1997) based on the assumption that fully molecular cores were needed to explain the dust absorption features and $H_2$ emission known at that time. Combining our results with Fesen et al.(1997)'s estimate of 3.4 M_sun in the form of ionized and previously-known neutral gas, the $H^0/H_2$ cores contribute about 5% of the total mass in the system of filaments.

In Paper II we noted very weak $H_2$ emission coming from many parts of the filaments outside of the catalogued $H_2$ knots, which could trace additional similar cores that do not efficiently emit $H_2$ because they do not have the optimum temperature and density for $H_2$ emission. These would not be counted in the above mass estimate. Figure 8 suggests that such cores would emit low surface-brightness [O I] and [S II] lines, but we have not run specific models of them.

---

[7] The predicted 21 cm emission from Knot 51 would only be $\sim 10^{17}$ cm$^{-2}$ K$^{-1}$, which is much too faint to be detectable with an instrument like the EVLA.

There could also be additional cores that in fact *are* fully molecular. We ruled out such a possibility for the particular case of K51 because the $H_2$ spectrum predicted by our dense core model does not match the observations. But there could be other knots for which we do not have a NIR spectrum that in fact would be described by our dense core model, or there could be cores that are too cold to strongly emit $H_2$ but which could be detected through CO observations.

**4.7 Crucial observational questions**

Further progress on understanding the $H_2$–emitting knots rests in part on filling in some missing key pieces of information. Here we describe four open questions that can be addressed in a practical way by new observations.

Where is the dust and what is the dust-to-gas ratio? As is discussed above, many $H_2$-emitting knots do not show accompanying dust absorption. Are they dust free? Dust absorption is seen in many locations where we have not been able to detect strong $H_2$ emission. What are the detailed correlations between the absorbing dust features and the $H_2$ emitting cores? We are currently using archival HST images to investigate this question, but the available continuum images do not fully sample HST's point spread function and have useful signal:noise ratio over only about 2/3 of the Crab. A survey to uniform depth over the full Crab at the higher spatial resolution offered by WFC3 would be very helpful.

What is the geometry? The existing $H_2$ images have only about 0.7" FWHM spatial resolution, which is insufficient to discern the exact relationship between the ionized and molecular regions. Observing the 2.12 μm line in K51 and a few other knots with an adaptive optics system on an 8 m-class telescope could produce images with spatial resolution similar to or better than the HST visible-wavelength narrow-band images, which would be a major improvement.

Does significant Hβ emission come from the same zone that emits $H_2$? Both of our models with extra heating say that it does. The available HST Hβ image (Fig. 2) is not nearly good enough to answer this question. A new HST Hβ image with high signal:noise ratio would directly test these models, which predict that the Hβ emission should largely trace the $H_2$ morphology, rather than coming from a surrounding sheath. It is important that K51 has a radial velocity that is measured to be close to zero, because this means that its emission lines fall within the velocity range covered by the HST narrow-band filters.

Finally, are there any fully molecular cores? CO observations are needed to search for these. Conversely, detection of strong CO emission from K51 would rule out our extra-heating models (§4.4). We are currently carrying out a program using the IRAM 30 m and Plateau de Bure telescopes for a preliminary search for CO emission from the Crab, but ALMA will be able to carry out a much deeper survey.

## 5. Conclusions

We have explored four different types of models of the regions in small knots within the Crab Nebula. Our goal is to simultaneously reproduce the classical optical emission lines and the exceptionally strong emission from $H_2$ that has a thermal distribution of populations. The first two of these models are powered only by the synchrotron radiation field and can be ruled out. This includes the pressure-balance model, which does not produce sufficiently strong $H_2$ emission from within the size scale of Knot 51, and the dense-core model, which can produce the necessary $H_2$ 2.12 μm emission by secondary electron excitation but predicts a highly non-thermal $H_2$ spectrum. From those results we deduced that extra heating is needed in the core of the knot in order to excite the observed $H_2$ emission.

We considered two cases for this additional heating: generic heating by any mechanism (such as shocks or dissipative MHD waves) that is able to supply a heating rate of $2.5 \times 10^{-16}$ ergs s$^{-1}$ cm$^{-3}$ (the temperature floor case), and as a specific heating mechanism, ionizing particles. As was discussed above, both of these models successfully reproduce the observed $H_2$ emission including both the surface brightness and the $H_2$ line strengths relative to each other. Both also successfully reproduce the forbidden lines from the more highly ionized ionized gas. The temperature floor case underpredicts the strengths of forbidden lines from neutral and low-ionization species while getting the recombination lines approximately right, while the ionizing particle case comes closer to correctly predicting the neutral and low-ionization forbidden lines but overpredicts the H I and He I recombination lines. The 5× dust model is able to also fit the observed $A_{V\text{-}pnt}$ to within a factor of two while still fitting the emission lines as well as do the other two models, but at the price of adding the dust abundance as an additional free parameter. We conclude that all three of these models fit the observations as well as the uncertain geometry allows.

We have found that the dust content in our simulations can be varied appreciably, from dust free to a factor of 5 above ISM abundances, while still maintaining overall reasonable fits to the ionized, neutral and molecular line strengths. Thus, we can conclude that dust might play an important role in understanding certain knots but it is not necessary for all knots.

The goal of this paper has been a general exploration of the types of models that might be able to explain the strong, thermal $H_2$ emission seen coming from the molecular knots. This paper presents a progress report, not a fully accurate model. In that spirit, we conclude that the models with extra heating do in fact represent progress and are likely to be a guide to what is really happening in the $H_2$-emitting zone. We have ruled out one of the two cases (dense cores) proposed by G90, while the second, ionizing particles, remains a possibility.

We have found an unusual astrophysical environment, independent of the exact $H_2$ excitation mechanism, in which the relatively large $n_e / n_H$ ratio allows $H_2$ to form most quickly through associative detachment rather than through grain catalysis. There are only a few published descriptions of such environments (e.g.Ferland et al. 1995), and this paper presents a different look at the possible conditions in supernova remnants.

We have assumed that steady state holds in all of our models. In §4.6 we showed that in order to accurately understand the expected emission, future simulations including time dependent chemistry must be performed. Time dependent plasma / chemistry simulations in an environment like the Crab filaments have never been performed and this paper provides the basis for such an investigation. The existence of $H_2$ requires a dense core to provide shielding. The observed $H_2$ temperature shows that this would be highly over-pressured relative to the envelope of ionized gas.

The mass contained in these extended $H^0/H_2$ zones is significant, amounting to about 0.1 M_sun for the sum of the 55 $H_2$-emitting knots reported in Paper II. This is about 5% of the total mass estimated to be in the form of ionized gas or in more conventional $H^0$ zones. The observed $H_2$ is producing the 2.12 μm line at essentially its peak emissivity per unit mass, so the Crab could easily include many additional extended $H^0/H_2$ zones or fully molecular zones that are too cool to produce detectable $H_2$ lines; CO observations are needed to look for such gas.

Improved high-angular resolution, high signal-to-noise images are needed, especially in the visible continuum, in Hβ and in the $H_2$ 2.12 μm line. These would better trace the morphological connections between the dust, the ionized gas, and the $H_2$. This would greatly improve our ability to discriminate between possible heating and $H_2$ excitation mechanisms. Finally, a survey of CO would provide the means to distinguish between models with dense molecular cores and our favored models, which include extra heating.

## ACKNOWLEDGEMENTS

CR, JAB and EL are grateful to NASA for support through ADAP grant NNX10AC93G. CR wishes to acknowledge the support of the Michigan State University High Performance Computing Center and the Institute for Cyber Enabled Research. GJF acknowledges support by NSF (0908877; 1108928; & 1109061), NASA (07-ATFP07-0124, 10-ATP10-0053, and 10-ADAP10-0073), JPL (RSA No 1430426), and STScI (HST-AR-12125.01, GO-12560, and HST-GO-12309). We thank the anonymous referee for some very helpful comments.

## REFERENCES.

Abdo A. A. et al., 2011, Sci, 331, 739

Atoyan A. M., Aharonian F. A., 1996, MNRAS, 278, 525

Baldwin J. A., Ferland G. J., Martin P. G., Corbin M. R., Cota S. A., Peterson B. M., Slettebak A., 1991, ApJ, 374, 580

Black J. H., van Dishoeck E. F., 1987, ApJ, 322, 412

Buehler R. et al., 2012, ApJ, 749, 26

Cazaux S., Tielens, A. G. G. M., 2002, ApJ, 575, 29

Dalgarno A., Yan M., Liu W., 1999, ApJS, 125, 237

Davidson, K., 1978, ApJ, 220, 177

Davidson, K., 1979, ApJ, 228, 179

Davidson K., Fesen R. A., 1985, ARA&A, 23, 119

Draine, B., 2003, ApJ, 598, 1017

Duyvendak J. J. L., 1942, PASP, 54, 91

Fabian A. C., Johnstone R. M., Sanders J. S., Conselice C. J., Crawford C. S., Gallagher J. S. III, Zweibel E., 2008, Nat, 454, 7207

Ferland G. J., 1978, ApJ, 219, 589

Ferland G., 2011, in Colpi M., Gallo L., Grupe D., Komossa S., Leighly K., Mathur S., eds, Proc. Sci., Narrow-Line Seyfert 1 Galaxies and Their Place in the Universe. SISSA, Trieste, PoS(NLS1)013

Ferland G.J., Fabian A.C., Johnstone R.M., 1994, MNRAS, 266, 399

Ferland G. J., Korista K. T., Verner D. A., Ferguson J. W., Kingdon J. B., Verner E. M., 1998, PASP, 110, 761

Ferland G. J., Fabian A. C., Hatch N. A., Johnstone R. M., Porter R. L., van Hoof P. A. M., Williams R. J. R., 2009, MNRAS, 392,1475

Fesen R. A., Kirshner R. P., 1982, ApJ, 258, 1

Fesen R., Blair W. P., 1990, ApJ, 351, L45

Fesen R. A., Shull J. M, Hurford, A. P., 1997, AJ, 113, 354

Gallant Y. A., Arons J., 1994, ApJ, 435, 230

Glaccum W., Harper D. A., Loewenstein R. F., Pernic R., Low F. J., 1982, BAAS, 14, 612

Glassgold A., Langer W., 1974, ApJ, 193, 73

Gomez, H. L. et al., 2012, preprint (arXiv:1209.5677)

Graham J. R., Wright G. S., Longmore A. J., 1990, ApJ, 352, 172 (G90)

Hennessy G. S. et al., 1992, ApJ, 395,L13

Henney W. J., Williams R. J. R., Ferland G. J., Shaw G., O'Dell C. R., 2007, ApJ, 671, L137

Henry R. B. C., MacAlpine, G. M., 1982, ApJ, 258, 11

Hester J. J., 2008, ARA&A, 46, 127

Hester J. J., Graham J. R., Beichman C. A., Gautier T. N. III, 1990, ApJ, 357, 539

Hoshino M., Arons J., Gallant Y. A., Langdon A. B., 1992, ApJ, 390, 454

Indriolo N., Geballe T. R., Oka T., McCall B. J., 2007, ApJ, 671, 1736

Indriolo N., Fields B. D., McCall B. J., 2009, ApJ, 694, 257

Le Bourlot J., Pineau des Forets G., Roueff E., 1995, A&A, 297, 251

Loh E. D., Baldwin J. A., Ferland G. J., 2010, ApJ, 716, L9 (Paper I)

Loh E. D., Baldwin J. A., Curtis Z. K., Ferland G. J., O'Dell C. R., Fabian A. C., Salomé

P., 2011, ApJS, 194, 30 (Paper II)

Loh E. D., Baldwin J. A., Ferland G. J., Curtis Z. K., Richardson C. T., Fabian A. C., Salomé P., 2012, MNRAS, 421, 789 (Paper III)

MacAlpine, G. M., Satterfield, T. J., 2008, AJ, 136, 2152

Marsden P. L., Gillett F. C., Jennings R. E., Emerson J. P., de Jong T., Olnon F. M., 1984, ApJ, 278, L29

Mayall N. U., Oort J. H., 1942, PASP, 54, 95

Miller J. S., 1973, ApJ, 180, L83

Mori K., Burrows D.N., Hester J.J., Pavlov G.G., Shibata S., Tsunemi H. 2,004, ApJ, 609, 186

Osterbrock D. E., Ferland G. J., 2006, Astrophysics of Gaseous Nebulae & Active Galactic Nuclei, 2nd edn. University Science Press, Mill Valley, CA   (AGN3)

Pequignot D., Dennefeld M., 1983, A&A, 120, 249

Sankrit R. et al., 1998, ApJ, 504, 344

Satterfield, T. J., Katz, A. M., Sibley, A. R., MacAlpine, G. M., Uomoto, A., 2012, AJ, 144, 27

Shaw G., Ferland G. J., Abel N. P., Stancil P. C., van Hoof P. A. M., 2005, ApJ, 624, 794

Sternberg. A., Neufeld, D. A., 1999, ApJ, 516, 371

Tavani M. et al., 2011, Sci 331, 736

Temim T. et al., 2006, AJ, 132, 1610

Temim T., Sonneborn G., Dwek E., Arendt R.G., Gehrz R., Slane P., Roellig T.L., 2012, preprint (arXiv:1205.2062)

Tielens A. G. G. M., Hollenbach D., 1985, ApJ, 291, 722

Trimble V., 1968, AJ, 73, 535

Trimble V., 1973, PASP, 85, 579

Verner E. M., Verner D. A., Korista K. T., Ferguson J. W., Hamann F., Ferland G. J., 1999, ApJS, 120, 101

Webber W.R. 1998, ApJ, 506, 329

Wu C. C., 1981, ApJ, 245, 581

Young J.S., Scoville, N.Z., 1991, ARA&A, 29, 581

**TABLES**

| Table 1. Archival HST Images ||||| 
|---|---|---|---|---|
| Line | Filter | No. Images | Total Expo. (s) | Dates |
| Hβ | F487N | 2 | 4000 | 1994 Feb 2 |
| [O III] 5007 | F502N | 2 | 5200 | 2000 Jan 26 |
| [O I] 6300 | F631N | 3 | 3900 | 2000 Jan 26 |
| [S II] 6720 | F673N | 2 | 2600 | 2000 Jan 26 |
| Continuum | F547M | 12 | 11600 | 1994 Mar – 1995 Dec |

| Table 2. Observed and Dereddened Emission Line Fluxes ||||
|---|---|---|---|
| Line | Observed $F$(Line)/$F$(H$\beta$) | Dereddened $I$(Line)/$I$(H$\beta$) | Estimated Uncertainty (%) |
| [O II] λ3727 | 7.38 | 12.48 | 25 |
| [Ne III] λ3869 | 1.59 | 2.55 | 10 |
| [Ne III] λ3967 | 0.38 | 0.58 | 50 |
| [Fe V] λ4071 | 1.34 | 1.96 | 50 |
| H I λ4340 | 0.42 | 0.54 | 25 |
| [O III] λ4363 | <0.25 | <0.32 | 50 |
| He I λ4471 | 0.16 | 0.19 | 50 |
| He II λ4686 | 0.28 | 0.31 | 25 |
| H I λ4861 | 1.00 | 1.00 | - |
| [O III] λ4959 | 2.45 | 2.35 | 5 |
| [O III] λ5007 | 7.31 | 6.87 | 5 |
| [Fe II] λ5159 | 0.08 | 0.07 | 50 |
| [N I] λ5198 | 0.17 | 0.15 | 25 |
| He I λ5876 | 0.65 | 0.46 | 25 |
| [O I] λ6300 | 5.06 | 3.29 | 10 |
| [O I] λ6363 | 1.78 | 1.14 | 10 |
| [N II] λ6548 | 3.06 | 1.88 | 20 |
| H I λ6563 | 4.82 | 2.96 | 10 |
| [N II] λ6584 | 9.35 | 5.71 | 10 |
| [S II] λ6716 | 8.62 | 5.10 | 10 |
| [S II] λ6731 | 11.25 | 6.64 | 10 |
| He I λ7065 | 0.25 | 0.14 | 50 |
| [Ar III] λ7136 | 1.27 | 0.69 | 25 |
| [Fe II] λ7155 | 0.26 | 0.14 | 50 |
| [O II] λ7320 | 0.35 | 0.18 | 50 |
| [O II] λ7330 | 0.28 | 0.15 | 50 |
| [Ni II] λ7377 | 1.24 | 0.63 | 25 |
| H$_2$ λ2.12 μm | 0.62 | 0.14 | 10 |
| Brγ λ2.17μm | 0.12 | 0.03 | 50 |
| Spitzer MIPS 24μm (21–26μm) | 57. | 11. | - |
| Herschel PACS blue (57–83μm) | 260. | 50. | - |
| Herschel PACS green (79–121μm) | 120. | 23. | - |
| Observed flux through full aperture: $F$(H$\beta$) = 1.04E-14 erg cm$^{-2}$ s$^{-1}$. Dereddened flux through full aperture: $I$(H$\beta$)=5.4E-14 erg cm$^{-2}$ s$^{-1}$. Dereddened surface brightness corrected for filling 15% of the 3.8"× 11.2" aperture: $S$(H$\beta$) = 8.4E-15 erg cm$^{-2}$ s$^{-1}$ arcsec$^{-2}$ ||||

| Element | Table 3. Chemical Abundances | | | |
|---|---|---|---|---|
| | Initial Abundances by Number[1] | Final Abundances by Number[1] | Final Abundances, Mass Fraction[2] | Final Mass Fraction[2] Relative to Solar |
| He | -0.16 | -0.53 | -0.27 | 0.28 |
| C  | -2.80 | -3.40 | -2.66 | 0.02 |
| N  | -3.84 | -4.25 | -3.45 | -0.37 |
| O  | -2.62 | -3.28 | -2.42 | -0.16 |
| Ne | -3.11 | -3.74 | -2.78 | 0.07 |
| Mg | -4.10 | -4.70 | -3.66 | -0.43 |
| Si | -     | -5.05 | -3.95 | -0.79 |
| S  | -4.38 | -4.71 | -3.55 | -0.16 |
| Cl | -     | -7.33 | -6.11 | -0.79 |
| Ar | -5.60 | -5.32 | -4.06 | 0.09 |
| Fe | -4.49 | -4.61 | -3.20 | -0.25 |

[1] Relative to H.
[2] Fraction of total mass.

| | Table 4 | | | | |
|---|---|---|---|---|---|
| | Model Parameters and Predicted/Observed Surface Brightness Ratios | | | | |
| | Case | | | | |
| Parameter | Constant Pressure | Dense Core (log $n_{core}$ = 6.0) | Temperature Floor | Ionizing Particles | 5x Dust |
| Log $\varphi(H)$ (cm$^{-2}$ s$^{-1}$) | 10.06 | 10.06 | 10.06 | 10.06 | 10.06 |
| log Thickness (cm) | 16.5 | 16.5 | 16.5 | 16.5 | 16.5 |
| Additional Heating | None | None | Temp. floor | $\xi_H / \xi_0 = 10^{5.3}$ | $\xi_H / \xi_0 = 10^{5.3}$ |
| log $n_{core}$ (H atoms cm$^{-3}$) | 3.6 | 6.0 | 5.1 | 5.25 | 5.25 |
| $T_{core}$ ($T$ in last zone) | 6890 | 108 | 2837 | 2890 | 2880 |
| Predicted $A_{V\text{-}pnt}$ | 0.1 | 1.7 | 0.3 | 0.4 | 2.0 |
| log $N_{H2}$ (cm$^{-2}$) | 10.6 | 20.5 | 16.9 | 17.0 | 17.2 |
| $<n(H^+)/n(H_{tot})>$ | 5.8x10$^{-1}$ | 1.9x10$^{-2}$ | 1.0x10$^{-1}$ | 1.0x10$^{-1}$ | 1.0x10$^{-1}$ |
| $<n(H^0)/n(H_{tot})>$ | 4.2x10$^{-1}$ | 5.6x10$^{-1}$ | 9.0x10$^{-1}$ | 9.0x10$^{-1}$ | 9.0x10$^{-1}$ |
| $<n(H_2)/n(H_{tot})>$ | 1.7x10$^{-9}$ | 4.2x10$^{-1}$ | 6.3x10$^{-4}$ | 5.5x10$^{-4}$ | 8.6x10$^{-4}$ |
| $n_e/n_H$ in core | 1.0x10$^{-1}$ | 7.8x10$^{-5}$ | 5.7x10$^{-3}$ | 1.6x10$^{-2}$ | 1.38x10$^{-2}$ |
| Line | Predicted/Observed Surface Brightness | | | | |
| H$_2$ $\lambda$2.12 μm | 0.0 | 0.9 | 0.9 | 1.1 | 1.5 |
| [O II] $\lambda$3727 | 0.7 | 0.6 | 0.7 | 1.0 | 0.8 |
| [Ne III] $\lambda$3869 | 0.8 | 0.6 | 0.7 | 1.1 | 0.9 |
| H I $\lambda$4340 | 0.8 | 0.8 | 0.9 | 2.0 | 1.3 |
| He I $\lambda$4471 | 0.7 | 0.7 | 0.8 | 1.2 | 0.9 |
| He II $\lambda$4686 | 1.3 | 1.0 | 1.2 | 1.2 | 1.0 |
| H I $\lambda$4861 | 1.0 | 1.0 | 1.1 | 2.3 | 1.6 |
| [O III] $\lambda$5007 | 0.9 | 0.7 | 0.8 | 1.2 | 0.9 |
| [N I] $\lambda$5198 | 0.4 | 0.2 | 0.3 | 1.6 | 1.3 |
| He I $\lambda$5876 | 0.9 | 0.9 | 1.0 | 1.6 | 1.2 |
| [O I] $\lambda$6300 | 0.2 | 0.1 | 0.2 | 0.7 | 0.5 |
| H I $\lambda$6563 | 1.0 | 1.0 | 1.1 | 2.5 | 1.8 |
| [N II] $\lambda$6584 | 0.4 | 0.3 | 0.3 | 0.5 | 0.4 |
| [S II] $\lambda$6716 | 0.3 | 0.2 | 0.3 | 0.8 | 0.7 |
| [S II] $\lambda$6731 | 0.3 | 0.2 | 0.3 | 0.9 | 0.7 |
| [He I] $\lambda$7065 | 0.8 | 0.8 | 0.8 | 1.3 | 1.0 |
| [Ar III] $\lambda$7136 | 0.8 | 0.7 | 0.8 | 1.0 | 0.8 |
| [Fe II] $\lambda$7155 | 0.6 | 0.5 | 0.6 | 2.0 | 1.5 |
| [O II] $\lambda$7320 | 1.2 | 1.2 | 1.2 | 2.3 | 2.2 |
| H I Br$\gamma$ $\lambda$2.17 μm | 0.9 | 1.1 | 1.1 | 2.1 | 1.7 |

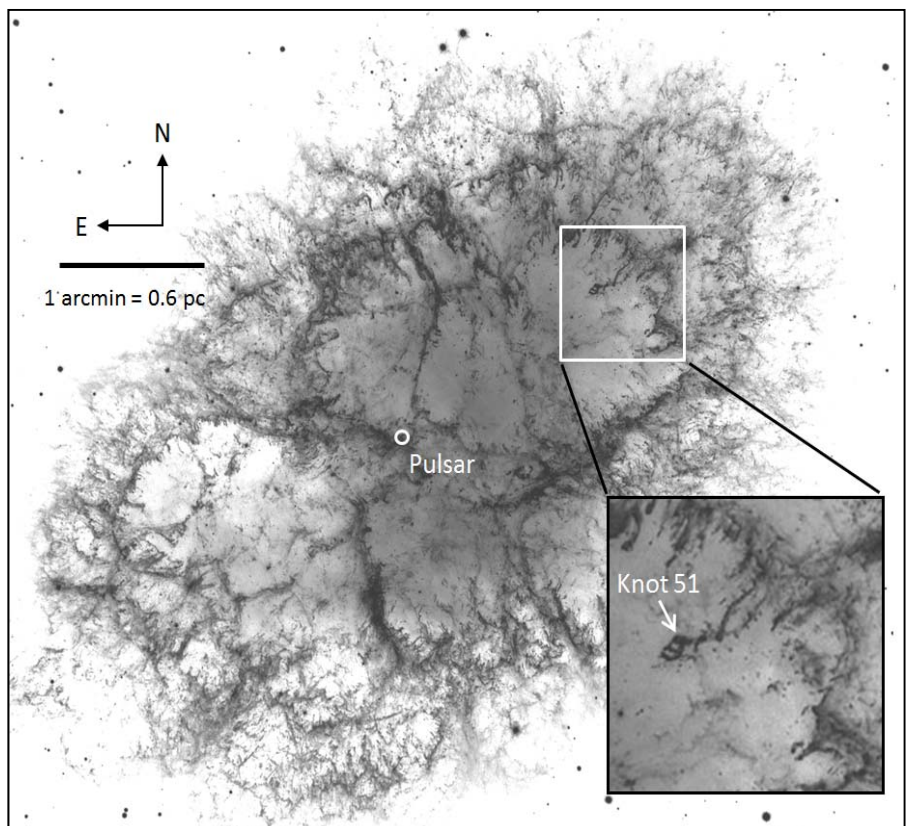

**Figure 1.** Positions of Knot 51 and the pulsar, shown on a grey-scale rendering of the well-known HST composite color image made with F502N, F631N and F673N filters (courtesy of NASA, ESA and J. Hester, Arizona State University; StScI News Release 2005-37).

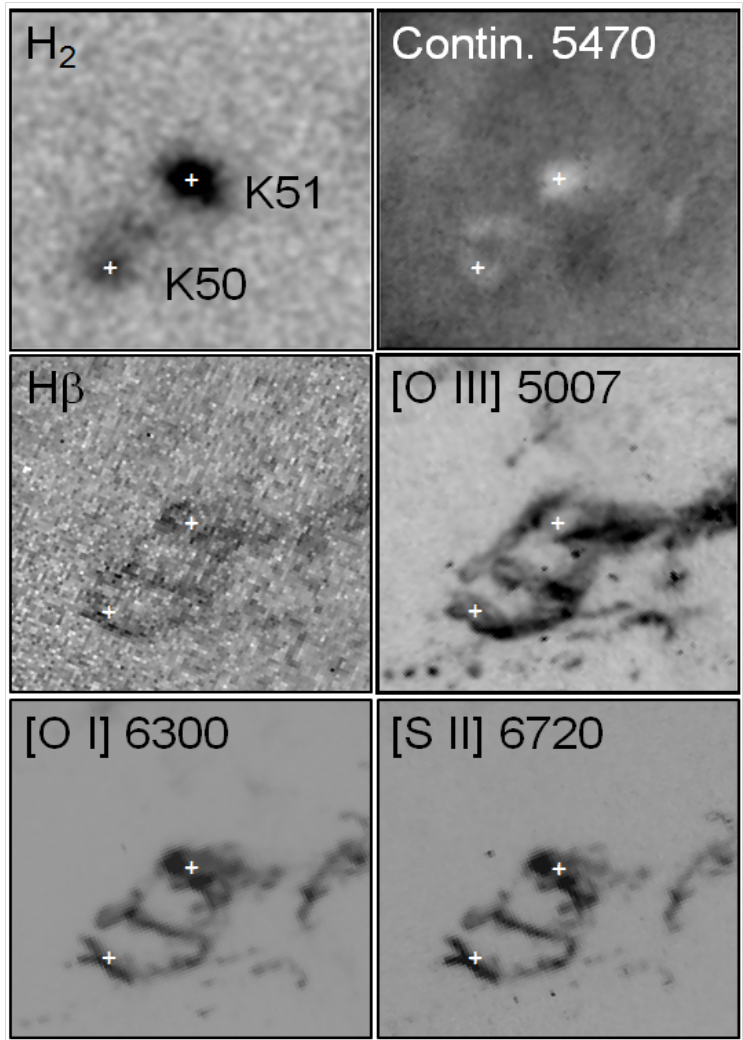

**Figure 2.** Images of Knots 50 and 51 through various filters. All boxes are 10"×10" and cover the same area on the sky with N up and E to the left. The pair of small white crosses on each image mark the position of the peak $H_2$ emission from the two knots. The $H_2$ image is from our ground-based imaging, the others are HST archival images. The [S II] $\lambda\lambda 6720$ and [O I] $\lambda 6300$ emission traces the $H_2$ 2.12 μm emission.

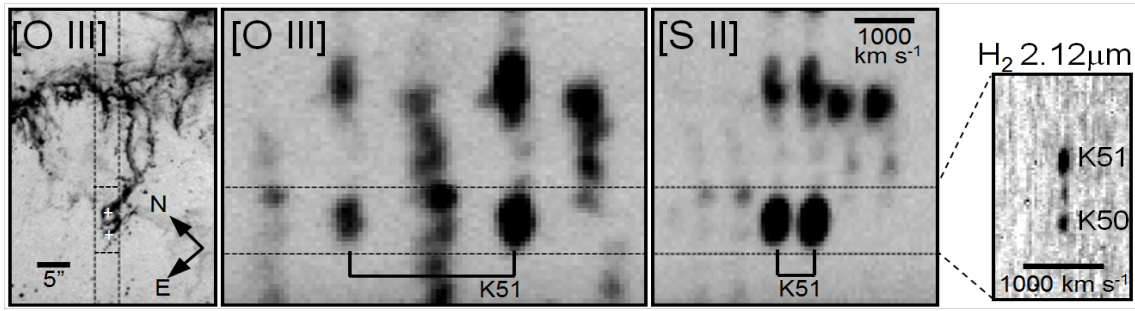

**Figure 3.** HST [O III] image (left panel) with vertical lines showing the area covered by our KPNO spectrum with its 3.8" wide slit. The two small white crosses mark the positions of Knots 50 (lower) and 51 (upper). Portions of the KPNO spectrum are shown in the middle two panels, aligned vertically with the HST image. The 11.2" region extracted along the slit is indicated by the horizontal lines, and the solid-lined goal posts mark the wavelengths of the [O III] $\lambda\lambda 4959, 5007$ and [S II] $\lambda\lambda 6716, 6731$ doublets from Knot 51 (blended along the slit with Knot 50). The velocity scale is the same in the two central panels, with velocity increasing to the right. Our NIR spectrum, taken through a 0.4" wide slit and with the scale on the sky magnified relative to the other images, is shown in the right panel. The higher spatial resolution of the NIR spectrum clearly separates Knots 50 and 51, and shows that they have the same radial velocity.

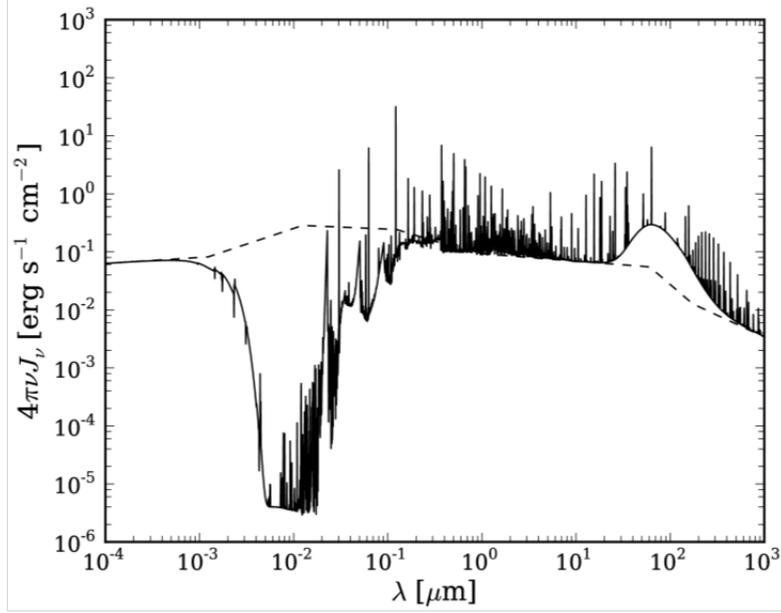

**Figure 4.** The spectral energy distribution in the center of the knot, including the transmitted synchrotron continuum, for the dense core model (solid line). The dashed line is the incident continuum. X-rays with energies $E = 1–10$ keV, corresponding to $\lambda = 10^{-3} – 10^{-4}$ μm, penetrate deep into the core of the knot and are preferentially absorbed by the inner shells of heavy elements. This produces high-energy Auger electrons and a shower of secondaries which then heat, ionize and excite the gas. This is the mechanism responsible for producing $H_2$ emission in the dense core model.

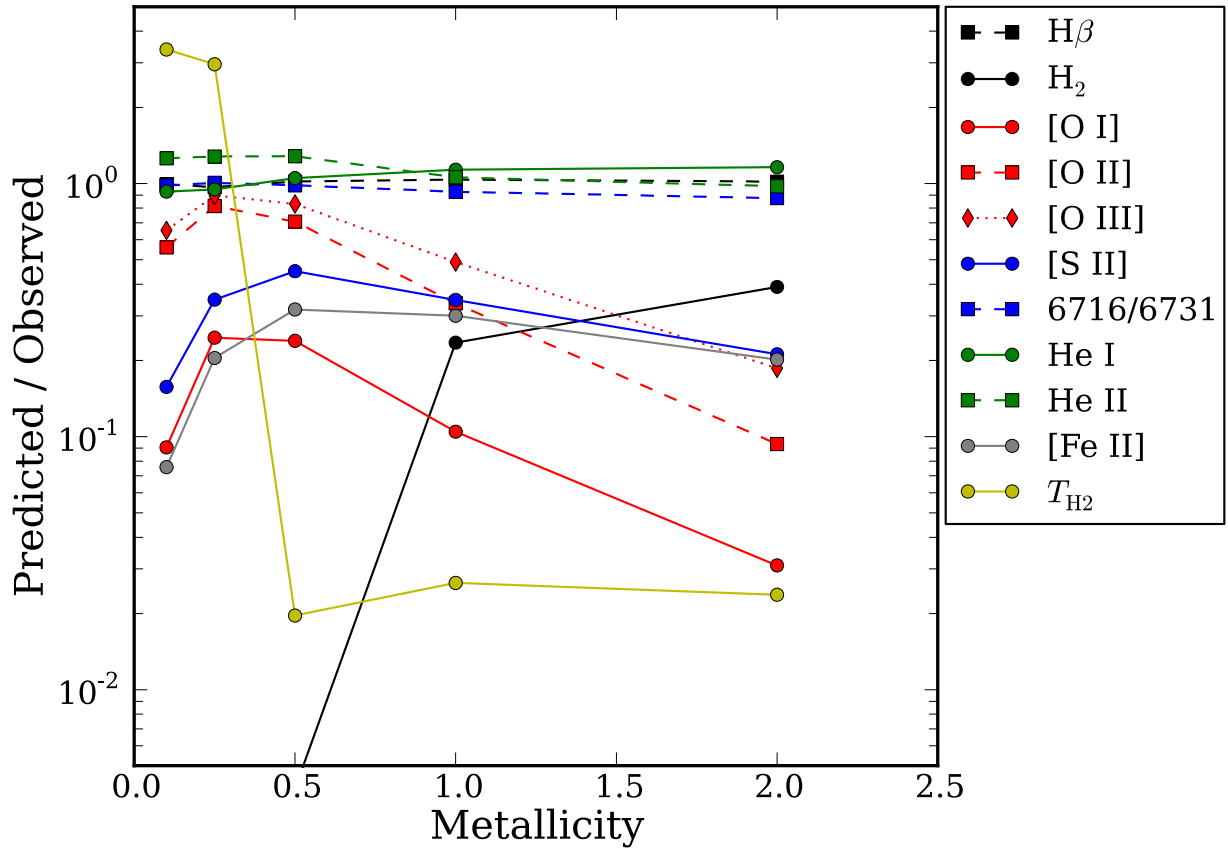

**Figure 5.** The effect of scaling metallicity in our constant pressure model, by number relative to the abundances obtained from our preliminary round of abundance optimization. As the metallicity decreases, the electron temperature rises and cooling shifts to the optical. The net effect is an increase in the optical forbidden line strengths. We adopt a scaling of 0.25, which gives the best fit to the higher ionization lines.

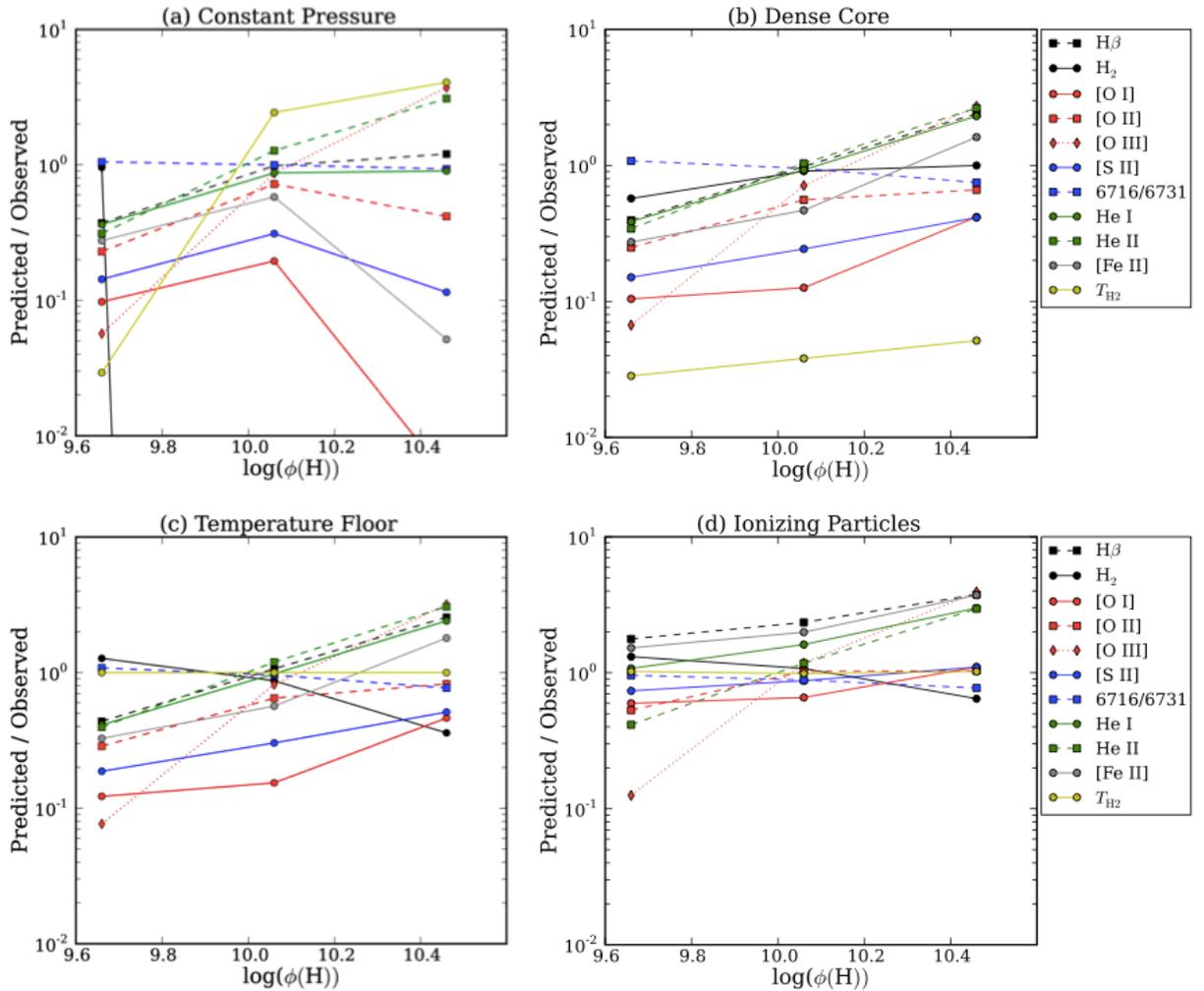

**Figure 6.** The predicted/observed ratios of surface brightness or intensity ratios of key emission lines and of $T_{H2}$, as a function of the ionizing flux $\varphi(H)$, for the constant-pressure case (a), dense-core case (b), temperature floor case (c), and ionizing particles case (d). In the ionizing particles case, H$\beta$ and He I depend only weakly on $\varphi(H)$ because ionizing particle heating accounts for over 50% of the surface brightness. In both cases the H$_2$ weighted core temperature is set almost entirely by the extra heating in the core rather than by the ionizing flux.

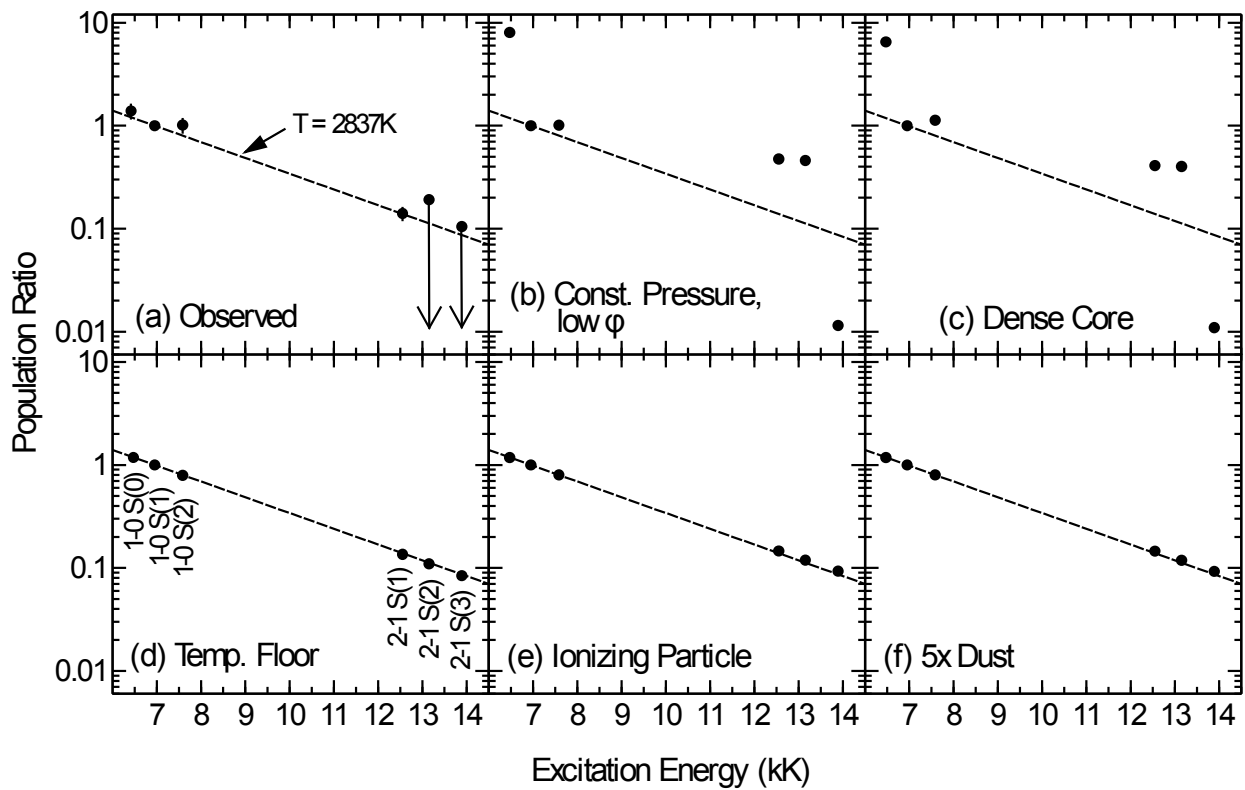

**Figure 7.** Observed and predicted $H_2$ level populations, normalized to the 1–0 S(1) 2.12 μm line. Panel (a): observed values for Knot 51 (from Paper III), showing four detected $H_2$ lines with error bars, and upper limits for two undetected lines. Panels (b), (c), (d), (e) and (f): predicted values for five models, showing that the constant-pressure model (here with $\log(\varphi(H)) = 9.66$) and dense core model (here with $\log(n_{core}) = 6.0$) do not fit the observations, while the temperature floor, ionizing particles, and 5x Dust models do fit.

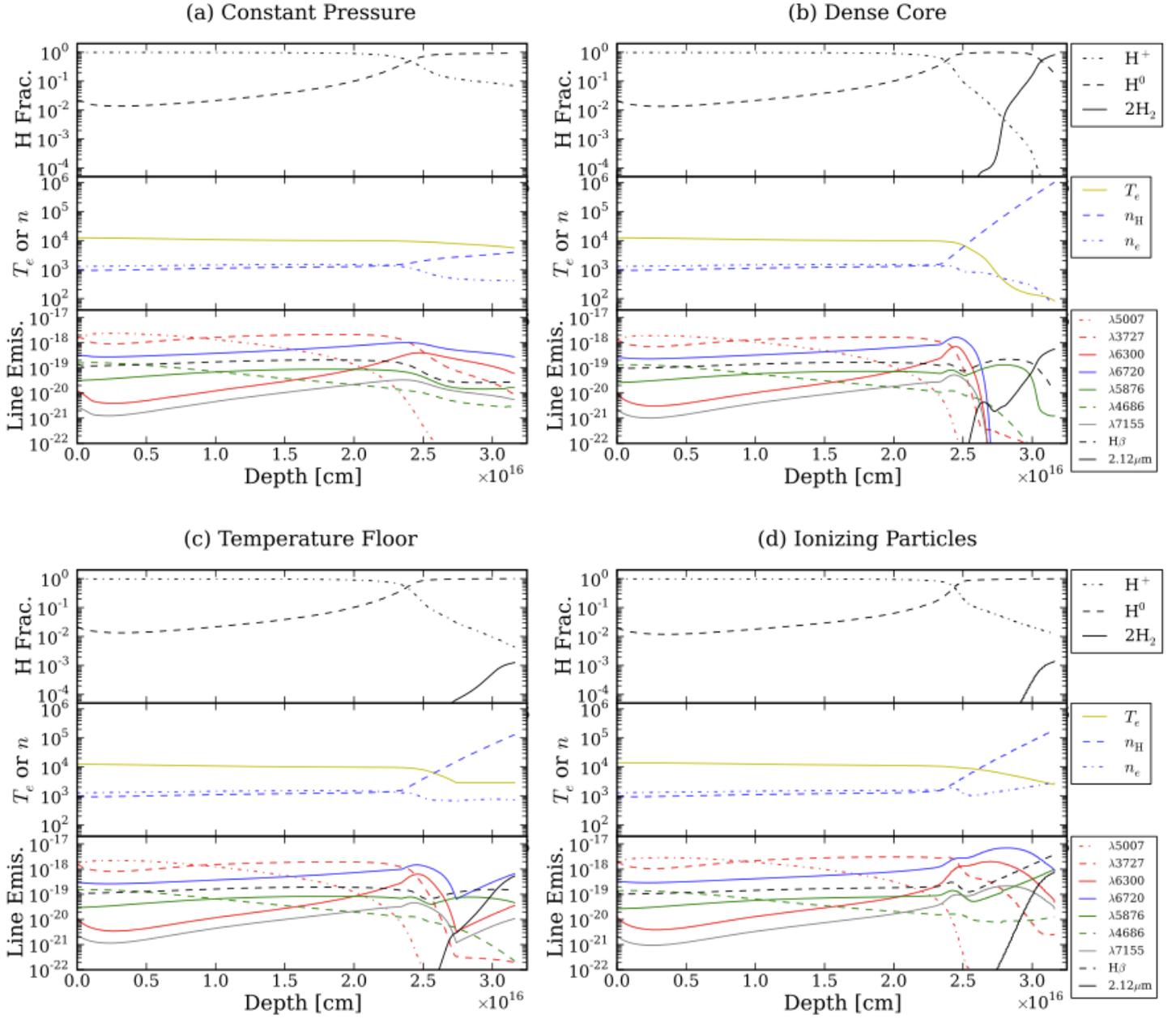

**Figure 8.** Each block of three panels shows the hydrogen ionization structure (top panel), $T_e$ [K] and densities [cm$^{-3}$] (middle panel), and line emissivities [erg s$^{-1}$ cm$^{-3}$] (bottom panel), as a function of depth. The different blocks show results for the constant pressure model (a), the dense core model (b), the temperature floor case (c), and the ionizing particles case (d). Comparison of the top panels in each block shows that of the three models which produce H$_2$ emission, the dense core model has a substantial molecular fraction in its center, while in the temperature floor and ionizing particles cases the H$_2$ is only a small trace constituent in an extended H$^0$ zone which emits H$_2$ as well as low-ionization and neutral forbidden lines.

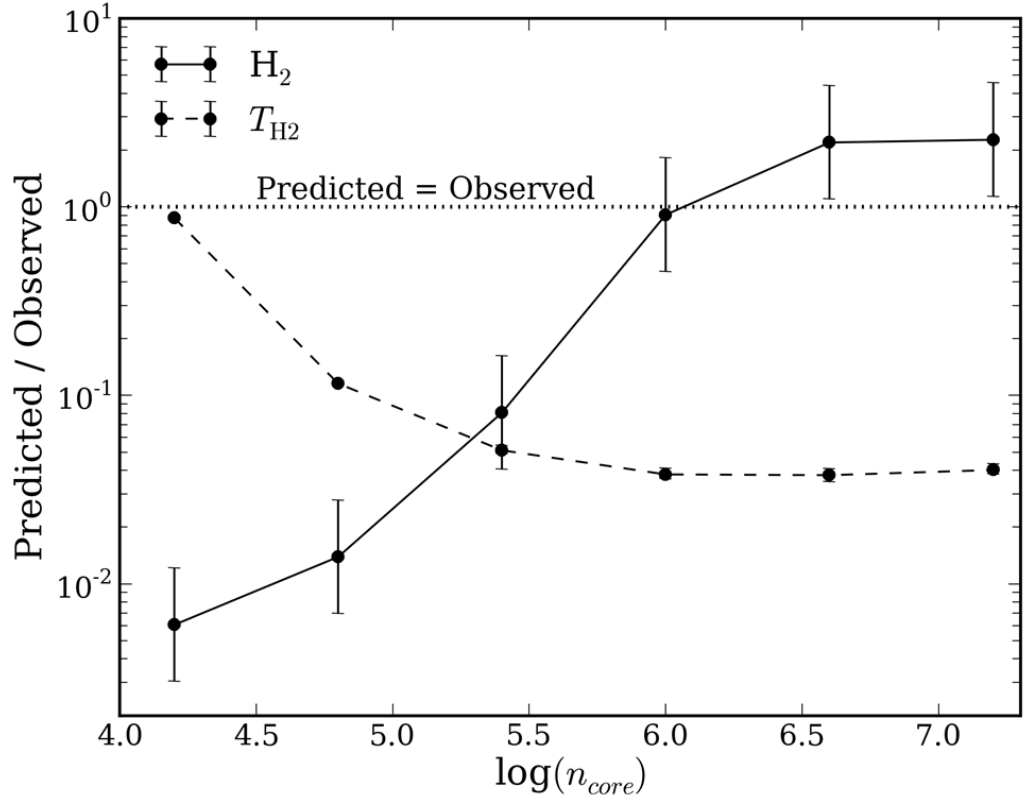

**Figure 9.** The predicted/observed ratio of the surface brightness of the 1-0 S(1) line of $H_2$ (solid line) and the predicted/observed temperature (dashed line), as a function of core density $n_{core}$, for the dense-core models. The factor-of-two error bars shown for the $H_2$ surface brightness reflect the uncertainty in the geometry of the knot. $T_{H2}$ is the $H_2$-weighted kinetic temperature. The model with the best-fitting $H_2$ surface brightness is for $\log(n_{core}) = 6.0$.

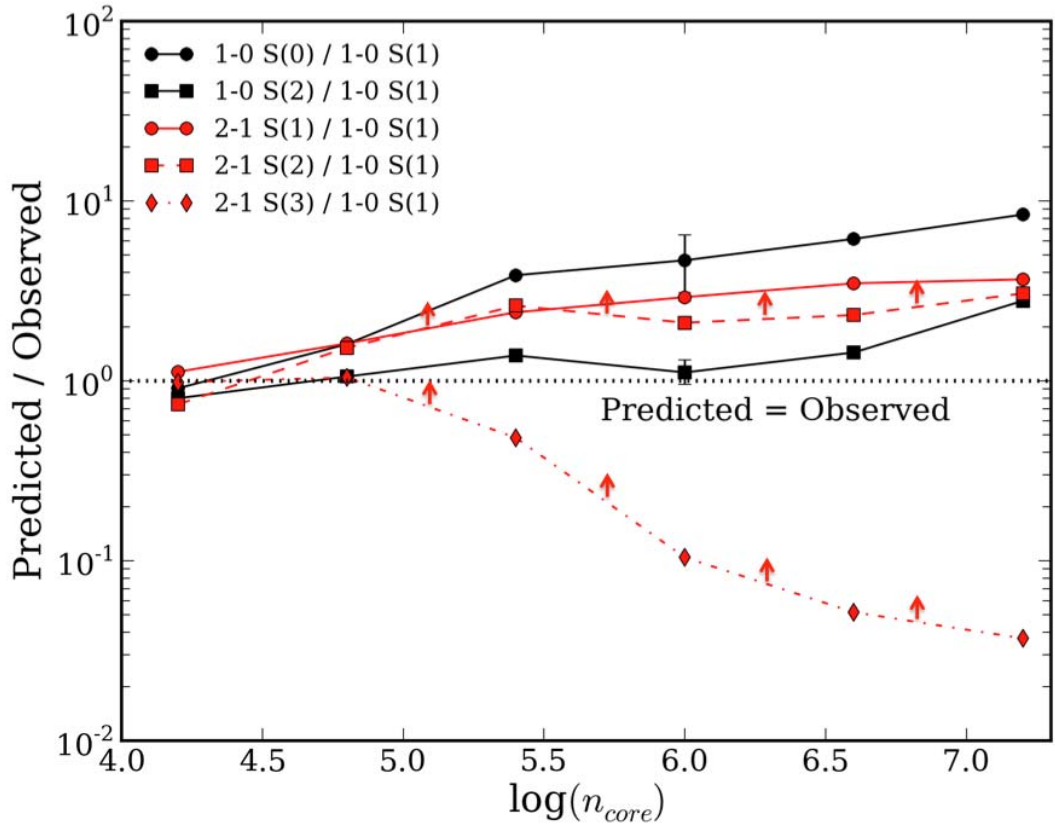

**Figure 10.** The predicted H$_2$ line ratios divided by the observed H$_2$ line ratios from Paper III, for the series of dense core models. In a perfect fit, all predicted/observed ratios would be 1. The two curves with upward arrows are cases where the observed value for the line in the numerator is only an upper limit, so that any predicted/observed ratio smaller than $10^0$ represents an acceptable fit. For each of the other curves, the error bars shown at a single point are the same size for all points. The model at log($n_{core}$) = 4.2, fits the observations the best but it is clearly not ideal, and Figure 9 shows that the 2.12 μm emission is too weak.

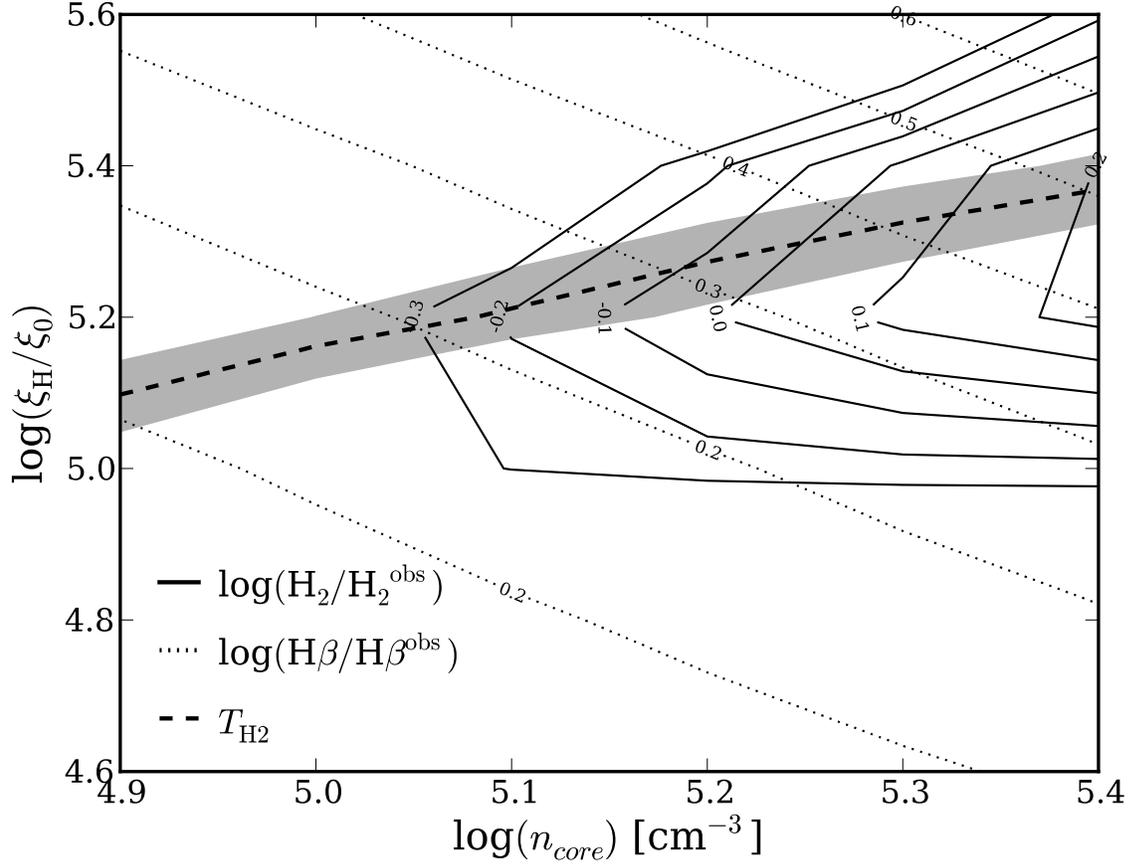

**Figure 11.** The optimization of core density and ionization rate scale factor for the ionizing particle case. Models reproducing the measured H$_2$ kinetic temperature fall along the dashed line while the gray band indicates the range due to the error bars of the measurement. The log contours show the ratio of the predicted to observed surface brightness of H$_2$ (solid line) and H$\beta$ (dotted line). We adopt $\xi_H / \xi_0 = 10^{5.3}$ and $n_{core} = 10^{5.25}$ cm$^{-3}$ as the optimal parameters despite the overestimated H$\beta$ emission.

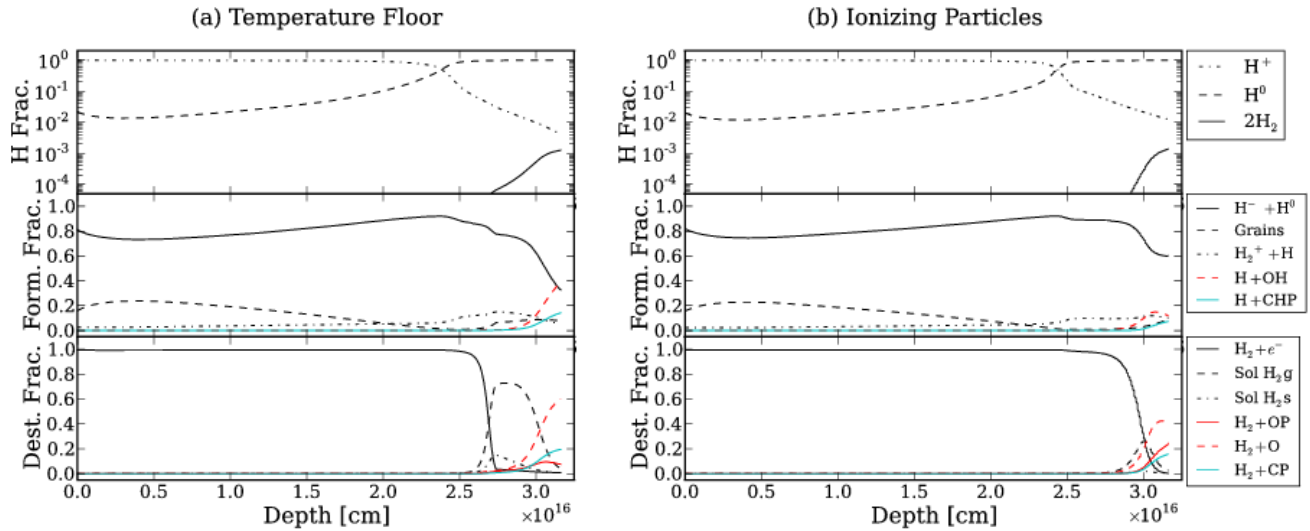

**Figure 12.** The $H_2$ formation mechanisms (middle panels) and $H_2$ destruction mechanisms (bottom panels), as a function of depth for the temperature floor case (a) and the ionizing particles case (b). A large fraction of $H_2$ forms through $H^-+H^0$ associative detachment (solid line in middle panels) rather than grain catalysis ("grains"). This is due to the relatively large $n_e / n_H$ ratio in the core (see middle panels, Figure 8). The hydrogen ionization structure is shown in the top panels for reference.

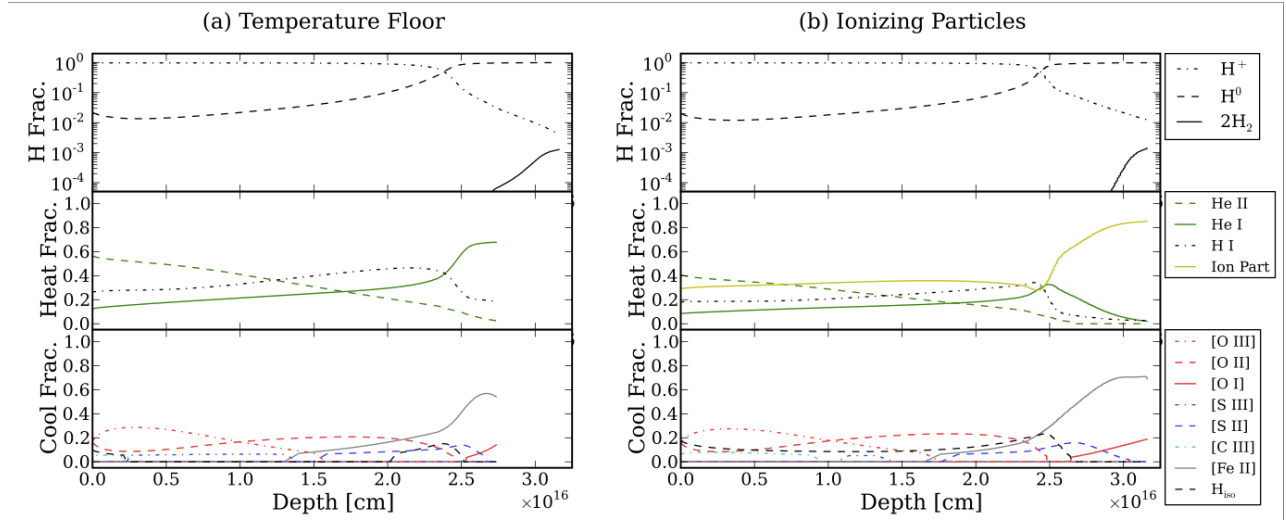

**Figure 13.** The heating fractions (middle panels) and cooling fractions (bottom panels) as a function of depth, for the temperature floor case (a) and the ionizing particles case (b). The hydrogen ionization structure is shown in the top panels for reference. In the temperature floor case, the knot is heated by photoionizations followed by recombinations of the different ions indicated in the legend. In the ionizing particles case, the knot is heated by the same processes as the temperature floor case in the $H^+$ zone, but heating by ionizing particles dominates in the $H^0/H_2$ zone. Cooling mainly proceeds through electron collisional excitation of [O III] in the ionized zone and Fe II in the core. For the temperature floor case, we do not show the heating and cooling in the deeper zones where the temperature is arbitrarily fixed at $T_e \approx 2800$ K.

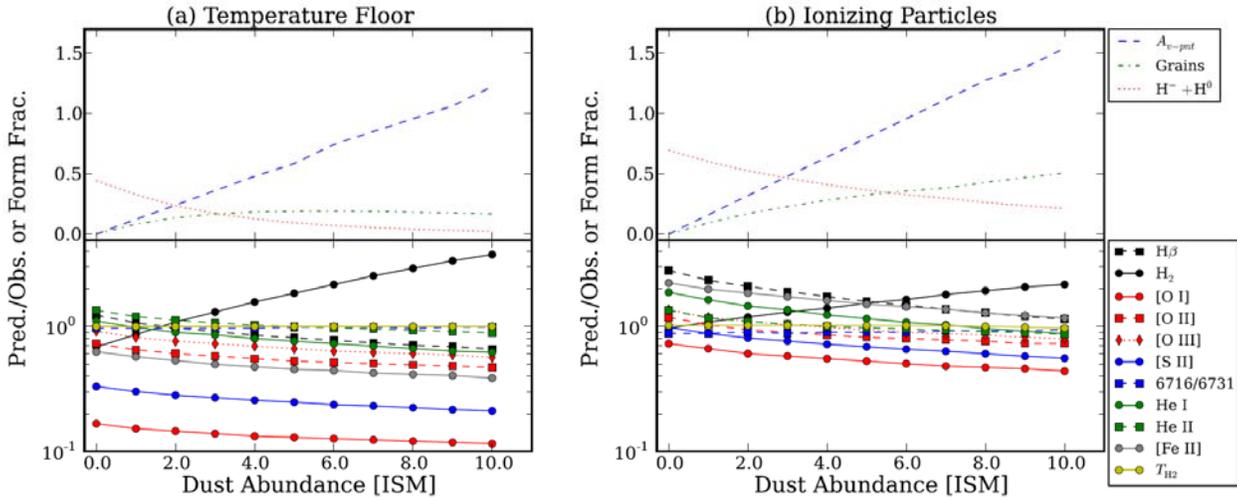

**Figure 14.** The sensitivity of several key parameters as a function of the grain abundance relative to the ISM grain abundance. The top panels show the formation fraction at the center of the knot due to the grain catalysis ("grains") and the $H^-+H^0$ formation routes and twice the predicted (scattering corrected) visual extinction $A_{V\text{-}pnt}$ relative to the observed values. The temperature floor case (a) is a better indicator of the true dependence on grain abundance than is the ionizing particles case (b), because in the temperature floor case the $H_2$ temperature is fixed at the observed value. Note that as the dust abundance becomes greater, the ionizing particle density has *not* been reoptimized to produce the correct $H_2$ temperature. At a grain abundance of ~3.0 times the ISM value, grain catalysis overtakes associative detachment in the temperature floor case. However, at those high grain abundances both of these $H_2$ formation fractions are small and $H+OH \rightarrow O+H_2$ (not shown) becomes dominant. The bottom panels show the predicted over observed values for various quantities as a function of grain abundance. The predictions of the model with a grain abundance of 5.0 times the ISM value are displayed in Table 4.

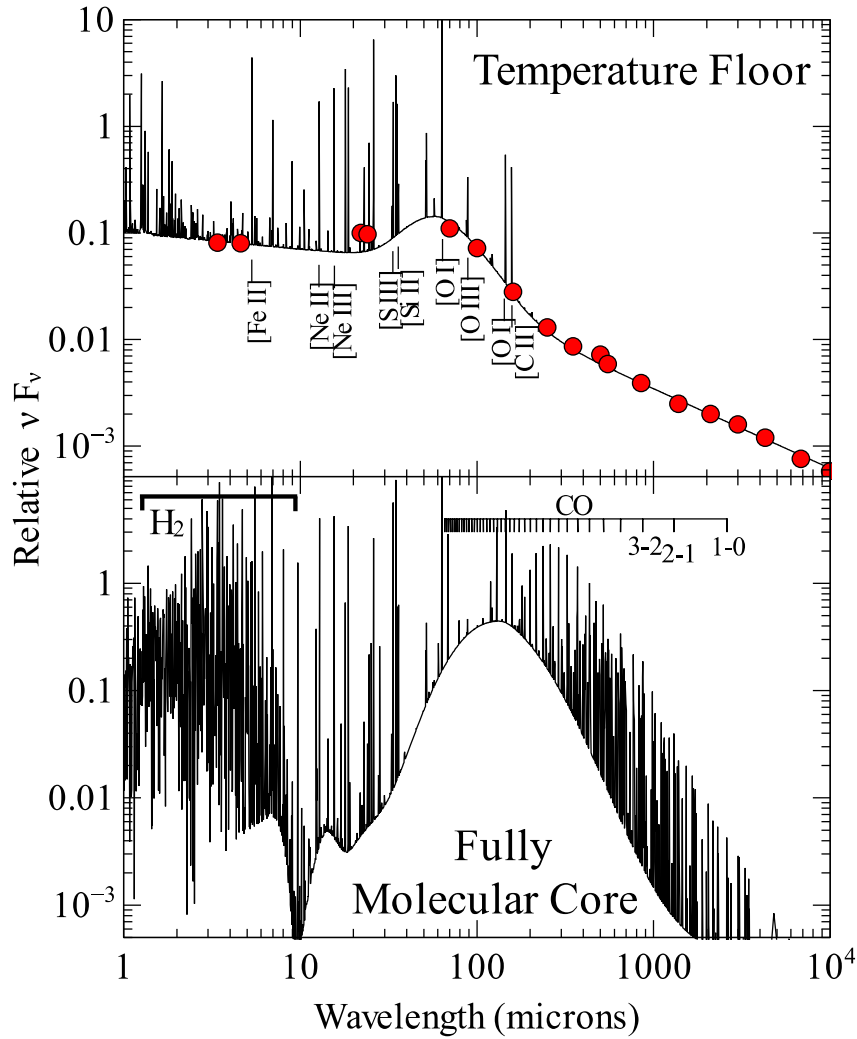

**Figure 15.** Predicted IR–mm wavelength spectra for the Temperature Floor model and for a typical fully molecular core, computed with CLOUDY. Examples of the stronger atomic lines are marked in the upper panel, while the CO v = 0 lines and the region dominated by $H_2$ emission lines are marked in the lower panel. The region longward of 100 μm in the molecular core model includes strong lines of CO, CS, HCN, SiO, $NH_3$, $H_2O$, SO and NO. The red dots show the continuum emission summed over a wide area on the Crab as measured by Gomez et al. (2012), after adjusting their observations to match the underlying continuum shape not due to dust emission and also the height of the dust emission feature predicted for a single $H_2$ knot. This shows that our model correctly predicts the peak wavelength and width of the dust emission feature.